\documentclass[aps,pre,nofootinbib,reprint,superscriptaddress]{revtex4-2}

\usepackage[utf8]{inputenc}
\usepackage{graphicx}
\usepackage{subfigure}
\usepackage{xcolor}
\usepackage{float}
\usepackage{placeins}

\usepackage{hyperref}

\newcommand{\be}{\begin{eqnarray}}
\newcommand{\ee}{\end{eqnarray}}

\newcommand{\wbe}{\begin{widetext}}
\newcommand{\wee}{\end{widetext}}

\usepackage{bbm}
\usepackage{amsmath,leftidx}
\usepackage{graphicx}
\usepackage{times}
\usepackage{CJK}
\usepackage{color,colortbl}
\usepackage[normalem]{ulem}

\def \bea {\begin{eqnarray}}
\def \eea {\end{eqnarray}}
\def \nn {\nonumber}
\newcommand{\eq}[1]{(\ref{#1})}

\begin{document}

\title{Work statistics for Quantum Spin Chains: characterizing quantum phase transitions, benchmarking time evolution, and examining passivity of quantum states}

\author{Feng-Li Lin}
\email{fengli.lin@gmail.com}
\affiliation{Department of Physics, \\
National Taiwan Normal University, Taipei, 11677, Taiwan}

\author{Ching-Yu Huang}
\email{cyhuangphy@thu.edu.tw}
\affiliation{Department of Applied Physics, \\
Tunghai University, Taichung 40704, Taiwan}

\begin{abstract}
We study three aspects of work statistics in the context of the fluctuation theorem for the quantum spin chains up to $1024$ sites by numerical methods based on matrix-product states (MPS). First, we use our numerical method to evaluate the moments/cumulants of work done by sudden quench process on the Ising or Haldane spin chains and study their behaviors across the quantum phase transitions. Our results show that, up to the fourth cumulant, the work statistics can indicate the quantum phase transition characterized by the local order parameters but barely for purely topological phase transitions. Second, we propose to use the fluctuation theorem, such as Jarzynski's equality, which relates the real-time correlator to the ratio of the thermal partition functions, as a benchmark indicator for the numerical real-time evolving methods.  Third, we study the passivity of ground and thermal states of quantum spin chains under some cyclic impulse processes. We show that the passivity of thermal states and ground states under the hermitian actions are ensured by the second laws and variational principles, respectively, and also verify it by numerical calculations. Besides, we also consider the passivity of ground states under non-hermitian actions, for which the variational principle cannot be applied. Despite that, we find no violation of passivity from our numerical results for all the cases considered in the Ising and Haldane chains.  {Overall, we demonstrate that the work statistics for the sudden quench and impulse processes can be evaluated precisely by the numerical MPS method to characterize quantum phase transitions and examine the passivity of quantum states. We also propose to exploit the universality of the fluctuation theorem to benchmark the numerical real-time evolutions in an algorithm and model independent way. }

\end{abstract}

\date{\today}

\maketitle
\tableofcontents


\section{Introduction}

Motivated by the progress of quantum simulation in the cold atom experiments \cite{bloch2008many,georgescu2014quantum, gross2017quantum,monroe2021programmable,scholl2021quantum} and the theoretical understanding of the thermalization of isolated quantum systems \cite{srednicki:1995pt,popescu2005foundations,goldstein:2005aib}, the non-equilibrium dynamics of quantum systems have been studied extensively \cite{polkovnikov:2010yn,aoki2014nonequilibrium,altman2015universal,d2016quantum,alet2018many}. 
A system can be driven into nonequilibrium simply by introducing a time-dependent interaction or a time-dependent coupling constant, and the system can return to equilibrium after turning off the time dependence of the coupling.  
For a many-body system, obtaining the exact dynamical evolution is generally difficult. Given a numerical method of evaluating the dynamical evolution, finding a way of estimating the numerical accuracy becomes a challenging task, especially for a system without prior knowledge of the exact dynamics. 
One would expect that the numerical error accumulates as the system evolves. Therefore, a reliable real-time error estimator will help compare the numerical results with the experimental ones, such as the quantum simulation.

For nonequilibrium dynamics of an isolated quantum system, one would expect that the underlying microreversibility should manifest in some way, in contrast with the macroscopic second law of thermodynamics. 
Indeed, equality relations exist, collectively known as fluctuation theorem \cite{jarzynski1997nonequilibrium,kurchan1998fluctuation,crooks1999entropy,tasaki2000jarzynski,kurchan2000quantum,talkner2007fluctuation,albash2013fluctuation,gong2015jarzynski,alhambra2016second,aaberg2018fully}, for such manifestation. 
The work done in a nonequilibrium process is not a state variable and will depend on the path connecting the initial and final state. The fluctuation theorem relates the average work to the free energy difference between the initial and final states. 
To calculate the average work or the higher moments, one can construct the corresponding characteristic function (or generating function of work statistics), which can be rewritten as the real-time correlation function \cite{tasaki2000jarzynski,talkner2007fluctuation,fei2020nonequilibrium}.  This characteristic function of work takes a form similar to the out-of-time-ordered correlator (OTOC) in characterizing the quantum chaos \cite{maldacena2016bound,halpern2017jarzynski,halpern2018quasiprobability}. 
In general, one needs to adopt some numerical method to calculate this generating function or some OTOC-like quantities involving nonequilibrium dynamical evaluation, which may be spoiled by the accumulation of numerical errors.
On the other hand, the difference in free energy only depends on the initial and final states. It will require disproportionately less numerical effort than the one needed for the characteristic function. Exploiting the feature of such disproportionality in the numerical efforts on both sides of the fluctuation theorem, one can use it to monitor the accuracy of any numerical method for evaluating the real-time nonequilibrium dynamical evolution of a quantum system.  In recent works \cite{gu2022tensor}, a numerical method based on the matrix-product-state (MPS) approach \cite{MPS1995, MPS2004} is adopted to evaluate the characteristic function for Ising spin chains and verify the fluctuation theorem, see also \cite{popovic2021quantum} for using the similar method for evaluating the heat statistics of open systems. This method can also be generalized to 2-dimensional spin-lattices, called the tensor-network state (TNS) \cite{PEPS2007}. 
In this paper, we will adopt the MPS method to evaluate the characteristic function of other quantum spin chains and use the fluctuation theorem to benchmark the capability of these numerical time-evolving methods.

Another interesting aspect of work statistics is characterizing the (dynamical) quantum phase transitions (QPTs). This was firstly demonstrated in \cite{PhysRevLett.101.120603} for the Ising chain and later in \cite{Chenu:2018spm} for the random matrix models under a sudden quench process, i.e,  with a sudden change of the Hamiltonian, for which the work characteristic function is related to the Loschmidt echo. Moreover, the average work done on the isolated system measures the expectation value of the jump of the Hamiltonian across the quenching point with respect to the initial state. Then, a similar trick is applied to show the universal scaling behavior of the kink statistics for the dynamical quantum phase transition under the Kibble-Zurek mechanism (KZM) for the (Ising) spin chains \cite{PhysRevE.89.062103, delCampo:2018hpn, fei2020work, Gomez-Ruiz:2022ewd}, and for Kitave honeycomb model \cite{zhang2022work}. This approach to characterizing QPTs by work statistics is extended to the (exactly solvable) spin/fermion chain models under the non-quench processes \cite{PhysRevE.89.062103, PhysRevResearch.2.033167, PhysRevResearch.2.033279}.

It is also known that entanglement structures like entanglement entropy can also characterize QPTs by showing a discontinuity at critical points in many cases of QPTs, e.g., see \cite{PhysRevA.66.032110, PhysRevA.81.032304}. As both the entanglement entropy and the work statistics can be encoded as some combinations of multi-point correlations, this may explain why both can characterize QPTs through some singular multi-point correlations. This perspective has been explored in \cite{Zawadzki:2023rsj} for the fermionic Hubbard model with random impurities. It showed that the entanglement is minimized while the work average is maximized at the critical point, but the second moment of work vanishes.  In this paper, we will study this issue further for the Ising and Haldane chains. 

When considering the quantum phase transition, the initial state is the ground state. After the sudden quench, it becomes a linear combination of the various excited states of the new Hamiltonian close in energy. If both initial and final Hamiltonians are non-critical and gapped, then only the gapped excited states with their energies less than the average work done on the system can be induced. 
However, if the final Hamiltonian happens to be at the quantum critical point, it describes a gapless system, then an infinite number of gapless excitations will be induced. 
We expect the work statistics to differ from those for the non-critical cases because this large number of gapless excitations will provide more dynamic paths for the quenching process for retrieving a definite amount of work. 
This implies that the average work excitation can display discontinuous behavior when the sudden quench connects a non-critical Hamiltonian to a critical one. In this sense, the work statistics, i.e., the average work done or the higher cumulants, can be adopted to characterize the quantum phase transitions, i.e., as an order parameter. Since the above reasoning concerns only the gapless feature of the critical systems, it can work for both the quantum phase transition of the Landau-Ginzburg type due to spontaneously-symmetry-breaking (SSB) or the topological type characterized by nonlocal order parameters. Finding the appropriate order parameter to indicate the topological phase transition is usually more difficult.  In this paper, we will explore the power of the numerical MPS-based algorithm to evaluate the moments/cumulants of work statistics and demonstrate their capability and limitations in characterizing the QPTs of the quantum Ising and Haldane chains. With the help of the quantum-state RG of MPS, we can evaluate the work statistics for the chains up to $1024$ sites. This will effectively suppress the finite-size effect, especially when near the quantum critical point.

Finally, we would like to study the issue of the passivity of ground states of quantum spin chains under cyclic impulse processes. It had been known that the relativistic thermal states are always passive, i.e., no work can be extracted from the system in any cyclic process \cite{Pusz:1977hb,1978JSP....19..575L,goldstein2013second}. We first show that the passivity of thermal states is guaranteed by the fluctuation theorem or, equivalently, the second law of thermodynamics. 
It is also straightforward to see that the passivity of ground states under hermitian action is guaranteed by the variational principle, i.e., the ground state has the lowest energy. However, the variational principle fails to ensure the passivity of the ground states if the action for driving nonequilibrium is non-hermitian. We then adopt the MPS numerical method to examine the passivity for such cases. In all cases we checked, we found no active ground states even under non-hermitian action. Besides, the patterns of average work extraction show some interesting features.

Based on the work statistics and the fluctuation theorem, we will use reliable numerical methods to investigate the above three aspects of the quantum spin lattice models in this paper: characterizing the quantum phase transitions, benchmarking the accuracy of the numerical real-time evolution,  and examining the passivity. These studies enlarge the perspective of the work done and the fluctuation theorem.  {In summary, in this work we will demonstrate the power of the numerical MPS method in evaluating the work statistics precisely enough to characterize the quantum phase transitions and examine the passivity of quantum states, even for the non-Hermitian actions. Moreover, we will also show how to exploit the simplicity and universality of the fluctuation theorem to benchmark the numerical real-time evolutions for generic models and numerical algorithms.}

The rest of the paper is organized as follows. 
For completeness, in the next section, we will briefly review the basics of work statistics and the fluctuation theorem and elaborate on the theoretical frameworks of the three aspects of work statistics we will address in this paper.
In section \ref{sec:3}, we will review the numerical methods for evaluating the real-time correlators based on MPS and then apply them to evaluate the generating function of work statistics and the expectation values of physical observables. 
In section \ref{sec:4}, we apply the above numerical methods to Ising-like and Haldane-like spin chains. Our numerical results demonstrate that the work statistics from the sudden quench can be used to characterize the quantum phase transitions for all spin models.  
In section \ref{sec:5}, we will show the results of using Jarzynski's equality as the benchmark to gauge the numerical accuracy of the real-time evolving methods. 
In section \ref{passivity_sec}, we present our results of examining the passivity of the ground states of the quantum spin chains under both hermitian and non-hermitian actions.  
Finally, in \ref{sec:6}, we summarize our works and discuss the possible extensions. Some numerical consistency checks and supplements are presented in the Appendices.

\section{Work statistics and its applications}\label{sec:2}

\subsection{Work statistics and fluctuation theorem}

We first give a brief sketch of the basics of work statistics in the context of the fluctuation theorem, which provides an elegant framework for describing and characterizing thermal or quantum fluctuations in statistical mechanics.

Unlike some conservative quantities, work is not a state variable and will depend on the microscopic details/paths of the non-equilibrium process. In short, work is a random variable and is characterized by a distribution function defined for work statistics as follows:
\be\label{pw}  
p(W)=\sum_{a,b} \delta \Big(W-[E_b(t_f)-E_a(0)] \Big) \, p(b,t_f|a) \;p_a 
\ee
where $p(b,t_f|a)$ is the transition probability from the energy eigenstate $|a\rangle$ of energy $E_a(0)$ at time $t=0$ to another eigenstate $|b,t_f\rangle$ of energy $E_b(t_f)$.
This non-equilibrium process is driven by a time-dependent Hamiltonian $H(t)$ from an initial state of density matrix $\rho(0)=\sum_a p_a |a\rangle\langle a|$,  with 
$H(0) \,|a\rangle=E_a(0) \,|a\rangle$ 
and  
$H(t_f)\,|b,t_f\rangle=E_b(t_f)\,|b,t_f\rangle$. 
 Thus, we have 
\be  
p(b,t_f|a)=|\,\langle b,t_f|U(t_f)|a\rangle \,|^2 = |\, \langle a|U^{\dagger}(t_f)|b,t_f\rangle \,|^2
\ee 
with $U(t_f)=\mathcal{T}e^{-i\int_0^{t_f} H(t) dt}$. For simplicity, in this paper, we will only consider the cases with $H(t)=H_0+\lambda(t) V$ with time-independent $H_0$ and $V$,  so that $[H(t), H(t')]=0$ and the time-ordering symbol $\mathcal{T}$ in $U(t_f)$ can be omitted. Moreover, we will in general allow for non-hermitian $H(t)$ so that $U(t_f)$ can be non-unitary, i.e., $U^{\dagger}(t_f)\ne U^{-1}(t_f)$.  The last equality implies detailed balance or micro-reversibility. 
Similarly, one can define the work statistics from the "reverse process" with an "initial state" of density matrix $\rho(0)=\sum_m q_m |m, t_f\rangle\langle m, t_f|$ as follows:
\be\label{tpw}
\tilde{p}(-W)=\sum_{a,b} \delta \Big( -W+[E_b(0)-E_a(0)] \Big)  \, \tilde{p}(a|b,t_f) \, q_b
\ee
where the transition probability for the reverse process is $\tilde{p}(a|b,t_f)=| \, \langle a|U^{\dagger}(t_f)|b,t_f\rangle \, |^2=p(b,t_f|a)$ with the last equality ensured by micro-reversibility.

The fluctuation theorem is usually formulated for the canonical initial and final states, i.e., $p_a=e^{-\beta(E_a(0)-F(0))}$ and $q_b=e^{-\beta(E_b(t_f)-F(t_f))}$ with $\beta={1\over k_B T}$ the inverse temperature, and $F(0)$ and $F(t_f)$ the free energy for the initial and final states, which are defined by the partition functions $Z(0)=e^{-\beta F(0)}$ and $Z(t_f)=e^{-\beta F(t_f)}$, respectively. 
Then, the fluctuation theorem taking the form of Crooks' relation \cite{crooks1999entropy} can be obtained by construction from Eq.~\eq{pw}-\eq{tpw}, and states 
\be  
p(W)\,e^{-\beta (W- \Delta F)}=\tilde{p}(-W)
\ee
where $\Delta F=F(t_f)-F(0)$. Integrate this relation over $W$, and we can obtain Jarzynski's equality \cite{jarzynski1997nonequilibrium, tasaki2000jarzynski}
\be 
\overline{e^{-\beta W}}=e^{-\beta \Delta F}
\ee
where the ''overline'' denotes the average over work $W$. Using Jensen's inequality, Jarzynski's equality can yield the second law: $\Delta S:= \overline{W}- \Delta F\ge 0$. The fluctuation theorem for more general quantum processes and end states can be found in \cite{albash2013fluctuation, alhambra2016fluctuating, aaberg2018fully}.

From the work statistics, we can define the corresponding  characteristic function as
\be
G(u)=\int dW \, e^{i u W} p(W)\;.
\ee
This function is the generating function of the moments of work done, e.g.,
\bea \label{w_avg}
\overline{W}&=&{1\over i}\lim_{u\rightarrow 0} {\partial \ln G(u) \over \partial u}\;, \\
\sigma^2_W &:=& \overline{W^2}- \overline{W}^2 =-\lim_{u\rightarrow 0} {\partial^2 \ln G(u) \over \partial u^2}\;.
\eea
For simplicity, later we will mostly adopt $\overline{W}=\lim_{s\rightarrow 0} {\partial G(-is) \over \partial s}$ since $G(-is)$ is real and $G(0)=1$. 

Moreover, Jarzynski's equality can be expressed as follows,
\be\label{Jarzynski_1}
G(i\beta)=e^{-\beta \Delta F}\;.
\ee

After some straightforward manipulation, the work characteristic function $G(u)$ of a non-equilibrium process, which drives the initial state $\rho(0)$ by a time-dependent Hamiltonian $H(t)$ during the time interval $[ 0, t_f ]$, can be put into the following form  \cite{talkner2007fluctuation},
\be \label{work_realtime}
G(u; t_f)= {{\rm Tr}[ \,U^{\dagger}(t_f)e^{i u H(t_f)} U(t_f) e^{-i u H(0)} \rho(0) \,] \over {\rm Tr}[ \, U^{\dagger}(t_f)  U(t_f) \rho(0) \,]} \;. 
\ee
The denominator is unity for unitary $U(t_f)$ (or hermitian $H(t)$).
The expression of Eq.~\eq{work_realtime} can be recast into the real-time correlation function on the extended Schwinger-Keldysh contour \cite{fei2020nonequilibrium} so that techniques of open system dynamics can be adopted. We usually omit $t_f$ and write $G(u)$.

To be specific, in this paper, we will only consider the time-dependent Hamiltonian of the following form,
\be\label{H(t)}
H(t)=H_0 + \lambda(t) V
\ee
where we choose the initial Hamiltonian $H_0=H(0^-)$ to be time-independent and hermitian; however, the driving operator $V$ is time-independent but could be non-hermitian. 
If $V$ is non-hermitian, then $U(t)$ is non-unitary so that the denominator of Eq.~\eq{work_realtime} is not unity. An initial state $\rho_0$, which is taken to be either the thermal state or ground state of $H_0$, will be driven away from equilibrium due to the nontrivial time-dependence of the coupling constant $\lambda(t)$. The notations $0^-$ and $0^+$ used later denote the moments right before and after $t=0$, respectively.

We will consider two particular non-equilibrium processes in the following: 
\begin{itemize}
\item Sudden quench process with 
\be \label{couple_sq}
\lambda (t) = \Delta \lambda \Theta (t)
\ee 
where $\Theta(t)$ is the Heaviside step function, and $\Delta \lambda$ a constant. For such a process, $U(t_f)\big \vert_{t_f\rightarrow 0}=1$, $H(0^-)=H_0$ and $H(0^+)=H_0 + \Delta \lambda V$. 

\item Impulse process with 
\be\label{couple_impulse}
\lambda(t) = \lambda \delta (t)
\ee
where $\delta(t)$ is the Dirac delta function and $\lambda$ a constant. Compared to a sudden quench process, the coupling constant is turned off right before and after the impulse so that $H(0^-)=H(0^+)=H_0$.
Thus, this process can be considered cyclic and will be implemented to study passivity. 
\end{itemize}

In both kinds of processes, we will find that the average work done $\overline{W}$ extracted from $G(u)$ can characterize (quantum) phase transitions.

\subsection{Benchmark the numerical real-time evolution}
 
Jarzynski's equality shown in Eq.~\eq{Jarzynski_1} is an interesting relationship that relates a real-time correlator to the ratio of the partition functions, i.e., $e^{-\beta \Delta F}={Z(t_f) \over Z(0)}$. There is usually no analytical method to calculate the real-time correlators such as $G(u)$ of Eq.~\eq{work_realtime} of a many-body system even though there might be some analytical ways of calculating the partition function. Even relying on the numerical method to evaluate the real-time correlators, the numerical errors accumulate more as the evolution continues. On the other hand, the partition function is a stationary quantity free of the accumulated error. Therefore, we can characterize the accumulated error by defining the following ratio:
\be
R(t_f)={G(i\beta) \over e^{-\beta \Delta F}}=G(i\beta) {Z(0)\over Z(t_f)} \label{benchmark_R}
\ee
which is the ratio of the LHS to the RHS of Jarzynski's equality and is a function of evolution time $t_f$.

By monitoring the deviation of $R(t_f)$ from the unity, one can estimate the numerical error accumulation of a numerical method for real-time evolution in a first-principle way without the need to compare with the analytical or other numerical methods. For different numerical methods for real-time evolution, we can compare the deviations of their $R(t_f)$ from the unity for different numerical methods for real-time evolution to benchmark their performances. Later, we will demonstrate benchmarking for the numerical methods adopted in this paper.

\subsection{Work done by sudden quench and phase transition}
\label{work done of sudden quench}

In general, the work statistics or its characteristic function for the many-body system is difficult to evaluate for the complication of real-time dynamics. However, to characterize the phase transition,  we can bypass the difficulty by just considering the work done by a sudden quench implemented by the Hamiltonian of Eq.~\eq{H(t)} and Eq.~\eq{couple_sq}. Denote the ground state of $H_0$ by $\rho_0 \equiv |0 \rangle\langle 0 |$, and the Hmailtoian after quench by $H_+$, i.e., $ H_+ \equiv H_0+ \Delta \lambda V$,  then the characteristic function can be simplified as follows:
\be 
\lim_{t_f \rightarrow 0^+} G(u) = {\rm Tr} \Big[e^{-i u H_0 } \rho_0 e^{i u H_+} \Big] 
 \label{G_sq}
\ee
which yields moments and the first few cumulants of the work done as follows, 
\bea \label{w_sq}
\overline{W^m}  &=& ( \Delta \lambda )^m \; \langle 0 |V^m | 0 \rangle \;, \\ \label{sw_sq} 
\sigma^2_W &=& \langle 0|H_+^2|0 \rangle -\langle 0 |H_+|0 \rangle^2 \;, \nn
\\
&=& (\Delta \lambda)^2 \Big( \langle 0 |V^2|0 \rangle -\langle 0|V|0\rangle^2 \Big)\;, \\
\kappa_3 &=& \overline{W^3} - 3 \overline{W^2} \overline{W} +  \overline{W}^3\;, \\
\kappa_4 &=& \overline{W^4} - 4 \overline{W^3} \overline{W} - 3 \overline{W^2}^2 + 12 \overline{W^2} \overline{W}^2 - 6 \overline{W}^4\;. \qquad 
\eea
The quantity $\overline{W}$, as shown for the sudden quench, measures the energy of the "excited state" $| 0 \rangle$ with respect to $H_+$, the quantity $\sigma_W$ measures the corresponding fluctuation of $\Delta H=\Delta \lambda V$. 
Thus,  $\overline{W}$ (or $\sigma^2_W$) could act as a local order parameter to characterize the quantum phase transition while tuning $H_0$. From Eq.~\eq{w_sq} we can also calculate the higher moments $\overline{W^m}$ or higher cumulants $\kappa_m$ for the sudden quench in terms of the expectation value of higher order $V^m$'s.

We now elaborate on why the quantities in Eq.~\eq{w_sq} and  Eq.~\eq{sw_sq} can be used as the order parameter to characterize the phase transition. First, suppose the initial and final states are Gibbs states related by unitary evolution. In that case, the Hamiltonian can also be understood as the modular Hamiltonian, i.e., $H_{\rm mod}=-\log \rho$. In this case, $\overline{W}$ can be rewritten as the relative entropy $S_{\rm rel}$ \cite{albash2013fluctuation}, i.e., 
\be 
\overline{W}={\rm Tr} [ \,\rho^{\rm G}_- \Delta H_{\rm mod}\,]=S_{\rm rel}(\rho^{\rm G}_-|\rho^{\rm G}_+) 
\ee
where $\rho^{\rm G}_{-, +}$ are the Gibbs states before and after quench, respectively. 
In this case, $\overline{W}$ measures the distance between the initial and final Gibbs states. For the case we consider, the initial and final states are pure, $\overline{W}$ is no longer equivalent to the relative entropy. However, as explained below, $\overline{W}$ can still be used to distinguish pure states, especially the critical and non-critical ones.

The phase transition is usually characterized by the order parameter, which is the expectation value of some local or non-local observables.  We can decompose the Hamiltonian $H_+=\sum_f E_{f^+} |f^+\rangle\langle f^+|$, so that Eq.~\eq{w_sq} with $m=1$ can be rewritten as 
\be \label{W_direct}
\overline{W}=\sum_f E_{f^+} |\langle 0 |f^+\rangle|^2 =\overline{E_{f^+}}\;.
\ee
The average work done $\overline{W}$ can be seen as the expectation value of $E_{f^+}$ with its probability measure given by $p(f)=|\langle 0 |f^+\rangle|^2$. We will tune some coupling in $H_0$ over a range covering different phases. 
Again, $|0 \rangle$ looks like the excited state to $H_+$. When $H_0$ and $H_+$ are in the same phase, we expect $|0^+\rangle \approx |0\rangle$ so that $p(f)=\delta_{f,0}$ so that  $\overline{W}\simeq E_{0^+} |\langle 0 |0^+ \rangle|^2 \simeq E_{0^+}$. On the other hand, if they are in different phases, we should not expect a sharp distribution for $p(f)$, and ${\overline W}$ will be quite different from $E_{0^+}$ due to the incoherent average. 
For example, when $H_+$ is the gapless critical Hamiltonian at the quantum critical point, we expect the emergence of a dense set of degenerate gapless states denoted as $\{ 0^+_c \}$ such that the density of ground states $p(0^+)=\sum_{\{ 0^+_c \}} p(c) < 1$, which is quite different from the case of non-critical gapped $H_+$ with $p(0^+)\simeq 1$. This then induces a sudden jump of $\overline{W}$ from $\overline{W}\simeq E_0^+$ to the incoherent average $\overline{W}\simeq   E_{0^+} \sum_{\{ 0^+_c \}} p(0^+_c) + \sum_{f\ne \{0^+\}} p(f) E_f$.
Therefore, $\overline{W}$ can be used as an order parameter for phase transition. Similar arguments go for $\sigma^2_W$ or higher moments/cumulants.  Some preliminary study of $p(f)$ for quantum Ising chain under sudden quench by the exact diagonalization (ED) method is done in Appendix \ref{app_a2}, and the result is shown in Fig. \ref{fig:pf_distribution} therein.

\subsection{Passivity of a quantum state: thermal or ground states} \label{sec:IID}

A quantum state is called passive if there is no work extraction, i.e., $\overline{W}\ge 0$, under a cyclic process defined by $H(t_f)=H(0)$. It was shown that the KMS state \cite{Kubo:1957mj, Martin:1959jp}, i.e., thermal vacuum state for a relativistic quantum field theory, is passive \cite{Pusz:1977hb,1978JSP....19..575L,goldstein2013second}. Even though there is no general criterion for checking the passivity of a generic quantum state, it can be argued from the fluctuation theorem as follows. For a cyclic process, we shall expect $\Delta F=0$ so that the fluctuation theorem becomes $\overline{e^{-\beta W}}\vert_{\rm cyclic}=1$. By Jensen's inequality, this implies a second-law statement
\be 
\Delta S = \overline{W}\vert_{\rm cyclic} \ge 0\;.  \label{2ndLaw_passive}
\ee 
Therefore, the fluctuation theorem guarantees the thermal states' passivity, thus, the second law. Since the second law is established only in the thermodynamic limit, there is evidence that the thermal-like states of finite systems under some peculiar process can be active \cite{allahverdyan2004maximal, skrzypczyk2014work, frey2014strong}. 

Based on the above result, it is worthy of mentioning the study in \cite{kaneko2019work}, which considered the passivity of the energy eigenstates of Ising chains under the action of local Hermitian operations. By adopting the eigenstate thermalization hypothesis \cite{Srednicki:1994mfb, Deutsch:1991msp, goldstein:2005aib}, an energy eigenstate locally looks like a thermal state, e.g., the reduced density matrix of a local region is approximately a thermal one with some effective temperature. Thus, based on the passivity of the thermal state and ETH, an energy eigenstate will probably be passive under the action of local unitary operations. In \cite{kaneko2019work}, it was shown that all the energy eigenstates are passive under local operations for non-integrable Ising chains of finite size, while most energy eigenstates are passive for integrable Ising chains. This is consistent with the fact that ETH holds only weakly for integrable systems, see, e.g., \cite{Lin:2016dxa, He:2017txy, He:2017vyf}. The above result could also help to explain the passivity of the ground states studied below and imply a negative effective temperature.

Intuitively, the ground states should be passive because the operation $V$ excites the ground state to those with larger energies so that one cannot extract energy from the ground state. We can check passivity by first obtaining the average work extraction $W_{\rm ext}^{\rm cyclic}\equiv -\overline{W}$  for the cyclic process with $H(t_f)=H(0):=H_0$ from the characteristic function of Eq.~\eq{work_realtime}, i.e.,
\be
W_{\rm ext}^{\rm cyclic} =
{\rm Tr} \big[  H_0 \rho_1 \big]- 
{\rm Tr} \big[  H_0 \rho_2 \big]  \;, \label{W_full}
\ee
where
\be
\rho_1 := {\rho_0 U^{\dagger}(t_f) U(t_f) \over {\rm Tr} \big[ \, \rho_0 U^{\dagger}(t_f) U(t_f) \,\big]}\;, \label{rho1}
\ee
\be
\rho_2 := {U(t_f)\rho_0 U^{\dagger}(t_f)  \over {\rm Tr} \big[ \, \rho_0 U^{\dagger}(t_f) U(t_f) \, \big]} \;. \label{rho2}
\ee
Note that ${\rm Tr}\rho_1 ={\rm Tr}\rho_2=1$\;.  Moreover, if $U(t_f)$ is unitary, i.e., $U^{\dagger}(t_f)=U^{-1}(t_f)$, then $\rho_1=\rho_0$ and $\rho_2=U(t_f)\rho_0 U^{\dagger}(t_f)$, so that Eq.~\eq{W_full} can be reduced to 
\be
W_{\rm ext}^{\rm cyclic} =
 {\rm Tr}  \big[ \,H_0 \rho_0 \,\big]- 
{\rm Tr} \big[ \, H_0 U(t_f) \rho_0 U^{\dagger}(t_f) \,\big]  \;, \label{W_full_h}
\ee
which is nothing but the energy difference between the ground state $\rho_0$ and the excited state $U(t_f) \rho_0 U^{\dagger}(t_f)$. Thus, in such cases, the passivity of the ground state is guaranteed if the variational principle holds. On the other hand, if $U(t_f)$ is non-unitary, then $\rho_1\ne \rho_0$ even though $\rho_2$ bears the look of excited state under non-unitary evolution. Therefore, the negativity of $W_{\rm ext}^{\rm cyclic}$ of Eq.~\eq{W_full} can no longer be ensured by the variational principle. 

We can conclude from the above discussions that the variational principle generally ensures the ground state's passivity. In this work, we will verify this by the numerical calculation and explore its possible violations by considering the cases of non-hermitian $V$ so that the variational principle cannot be applied to guarantee the negativity of $W_{\rm ext}^{\rm cyclic}$ of Eq.~\eq{W_full}.

For simplicity and numerical implementation, we will only consider the impulse cyclic processes, i.e., with $\lambda(t)$ given by Eq.~\eq{couple_impulse},  for which we have 
\be
\lim_{t_f\rightarrow 0^+} U(t_f)=\lim_{t_f\rightarrow 0^+} e^{-i\int_{0^-}^{t_f} dt [H_0+ \lambda \delta(t) V]]}=e^{-i \lambda V}\;.
\ee
This yields
\be
G(u)={{\rm Tr}\big[ \, e^{i\lambda V^{\dagger}} e^{i u H_0} e^{-i \lambda V}   e^{-i u H_0} \rho_0 \, \big] \over {\rm Tr} \big[ \, e^{i\lambda V^{\dagger}} e^{-i \lambda V} \rho_0 \,\big]}\;. \label{imp_G}
\ee
It turns out that we can easily implement MPS formalism for spin chain models to evaluate the right side of Eq.~\eq{imp_G} numerically and then extract $W_{\rm ext}^{\rm impulse}$ from the results. 
Alternatively, the average work extraction can also be obtained from Eq.~\eq{W_full}-Eq.~\eq{rho2} by replacing $U(t_f)$ with $e^{-i \lambda V}$ and $U^{\dagger}(t_f)$ with $e^{i \lambda V^{\dagger}}$.

It is easy to see that $W_{\rm ext}^{\rm impulse}$ vanishes exactly if $[V, H_0]=0$. Otherwise, by assuming $H_0|0\rangle=0$ and taking small $|\lambda|$ expansion up to ${\cal O}(\lambda^4)$, we have 
\bea
&& W_{\rm ext}^{\rm impulse} = -\lambda^2 \; \langle V^{\dagger} H_0 V \rangle \nn \\
&& +{i \over 2} \lambda^3 \Big[ \langle  V^{\dagger} H_0 V^2 - V^{\dagger 2} H_0 V \rangle
+ 2 \langle V^{\dagger} -V \rangle \langle V^{\dagger} H_0 V\rangle \Big] 
\nn \\
&& -{1\over 6} \lambda^4 \Big[ \langle {3\over 2} V^{\dagger 2} H_0 V^2 -  V^{\dagger 3} H_0 V - V^{\dagger} H_0 V^3 \rangle \nn \\
&& \;\; -3 \langle V^{\dagger} -V \rangle \langle  V^{\dagger} H_0 V^2 - V^{\dagger 2} H_0 V \rangle \nn \\
&& \;\; +3 \langle V^{\dagger} H_0 V \rangle \langle V^{\dagger 2}+ V^2 - 2 V^{\dagger} V \rangle  \Big]\;,
\eea
where we short-handed $\langle 0|\cdots|0\rangle$ into $\langle \cdots \rangle$. 
We see that the ${\cal O}(\lambda)$ term is absent, so that at ${\cal O}(\lambda^2)$, $W_{\rm ext}^{\rm impulse}$ is independent of the sign of $\lambda$. Note also that the last terms in each order shown above all vanish for $V^{\dagger}=V$.
Though this expression contains the terms with imaginary $i$ factor, they are always associated with the expectation values of anit-hermitian operators, so the overall expression is a real quantity. 

Besides, due to the peculiarity of the impulse process, there are some interesting features of $W_{\rm ext}^{\rm impulse}$. Firstly, if $V=\sum_k \sigma_i^a$ for the spin-$1/2$ chain with $\sigma_k^a$ for $a=1,2,3$ are the Pauli matrices for the spin operator at site $k$. Then, $e^{-i \lambda V}=\prod_k e^{-i \lambda \sigma_k^a}=\prod_k (\cos\lambda - i \sigma_k^a \sin\lambda)$, which is periodic in $\lambda$. The resultant $W_{\rm ext}^{\rm impulse}$ is then also periodic in $\lambda$.  Secondly, if $V= i\sum_k \sigma_i^a$ for the spin-$1/2$ chain, similar to but not the same as the hermitian case, $e^{-i \lambda V}=e^{i \lambda V^{\dagger}}= \prod_k e^{\lambda \sigma_k^a}=\prod_k (\cosh\lambda + \sigma_k^a \sinh\lambda)$. Using this to evaluate $W_{\rm ext}^{\rm impulse}$ and using the fact that $H_0|0\rangle=0$, we can obtain 
\be
\lim_{\lambda \rightarrow \infty} W_{\rm ext}^{\rm impulse} = - {\langle  \,\prod_k (1+ \sigma_k^a) H_0  \prod_j (1+ \sigma_j^a) \,\rangle \over 2 \langle \, \prod_k (1+ \sigma_k^a)\, \rangle}
\ee
which saturates to a negative finite constant as $\lambda \rightarrow \infty$. Note that the negativity is ensured by the variational principle and $|\langle \sigma_k^a \rangle|\le 1$. The peculiar features for both types of $V$ of spin-$1/2$ chains are due to the highly non-adiabatic nature of the impulse process. Similar conclusions can be drawn for the spin-$1$ chain with some specific $V$ composed of the sum of site-spin operators.

From the above discussion, we see that the passivity (or the negativity of $W_{\rm ext}^{\rm impulse}$) is guaranteed by the variational principle in the small $\lambda$ regime no matter if $V$ is hermitian or not. Therefore, when exploring the possibility of violating the passivity of the ground states for the non-hermitian $V$, we will investigate the large $\lambda$ regime, to which the variational principle cannot be directly applied. 

Finally, we comment on the possible physical realization of a non-Hermitian operation, although our consideration in this work is somewhat academic. There are ways to realize non-Hermitian actions. One way is to drop the quantum jump term in the Lindblad master equation for the open quantum system by choosing the proper Lindblad operator, and the resultant equation describes the Heisenberg evolution of the density matrix operator by a non-Hermitian Hamiltonian, see, e.g., \cite{roccati2022non}.  This can be alternatively realized on quantum circuits by probability quantum computing with suitably dilated Hilbert space, see \cite{schlimgen2022quantum}. Or, one can introduce the non-Hermitian interaction in the context of $\mathcal{PT}$-symmetry. For the Ising chain, it is equivalent to introducing an imaginary magnetic field; see, e.g., \cite{Castro-Alvaredo:2009xex}.

\section{Numerical methods for imaginary- and real-time evolution} \label{sec:3}

To calculate the work statistics, we must evaluate the associated generating function, i.e., the real-time correlator of the characteristic function, Eq.~\eq{work_realtime}. Even for a quenching process, it still involves a nontrivial vacuum expectation value (VEV) as given by Eq.~\eq{G_sq} or the average work done given by Eq.~\eq{w_sq}. 
This is usually a challenging task for many-body systems. To study the work statistics for various many-body quantum systems, a recent work \cite{gu2022tensor} has adopted the numerical method based on matrix-product-state (MPS) \cite{MPS1995, MPS2004} for quantum spin chains. This method is called time evolving black decimation (TEBD) \cite{TEBD2007,iTEBD2008} to use for the real-time evolution or evaluation of the VEV by its imaginary time evolution. Below, we will review the basic ideas of this method and mention its application for our purpose. 

Moreover, Jarzynski's equality Eq.~\eq{Jarzynski_1} relates a real-time correlator to the ratio of the partition functions of thermal states, which is far easier to evaluate than the former.  Therefore, we can use Jarzynski's equality to gauge the accuracy of the numerical real-time evolution and serve as the benchmark for a given numerical method for real-time evolution. In the next section, we will benchmark the TEBD and the exact diagonalization (ED).

\subsection{Matrix Product States} 
\begin{figure}
 \includegraphics[width=0.4\textwidth]{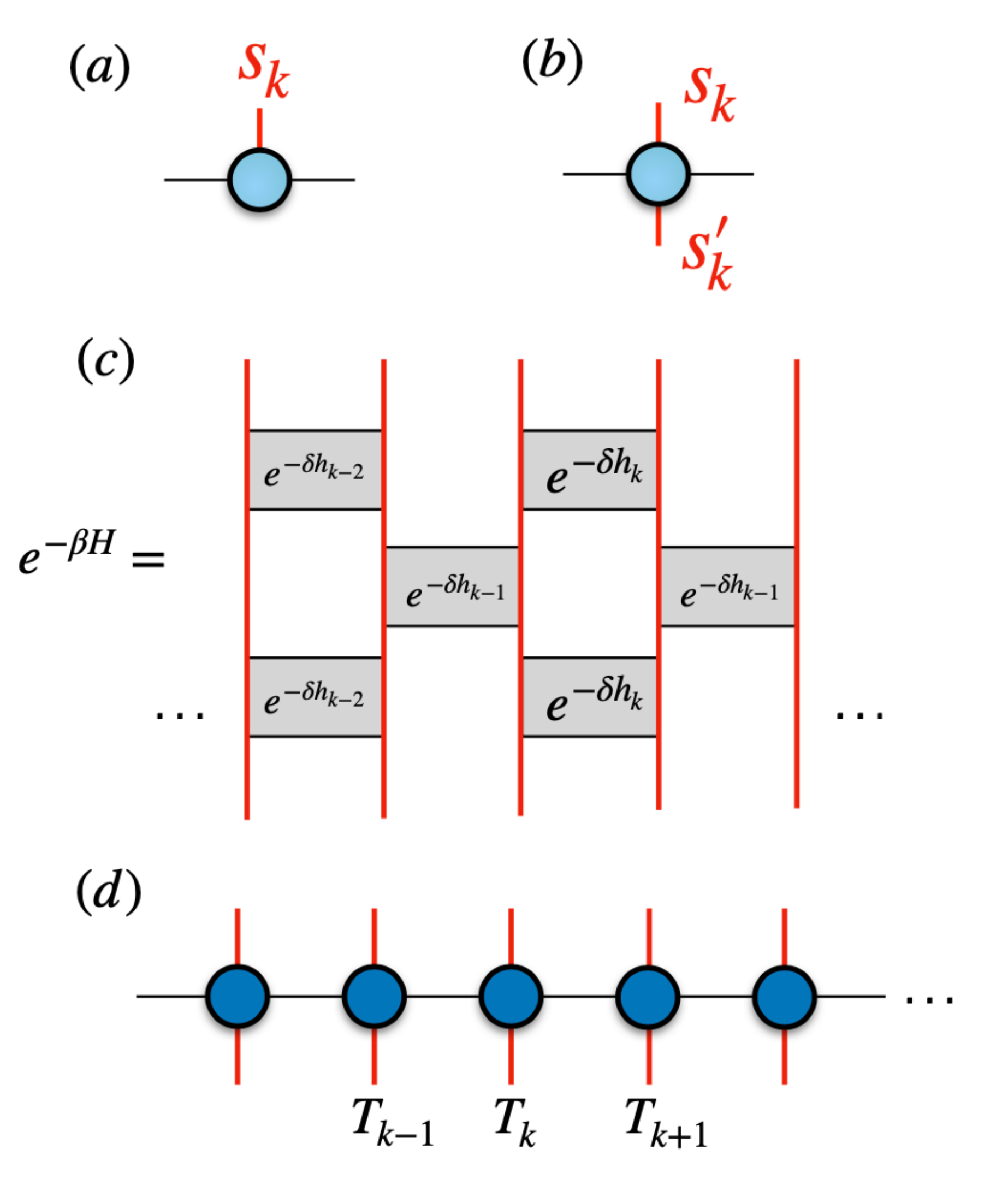}
 \caption{\small 
 (a) Symbol for matrix product state (MPS) structure.  
   (b)  Symbol for matrix product operator (MPO) structure. (c) MPO representation of the operator $ e^{-\beta H } $ for a quantum system with nearest-neighbor interaction via TEBD algorithm. 
(d) MPO representation of approximated   $ e^{-\beta H } $.   }
 \label{fig:rho_TN}
\end{figure}

For the one-dimensional quantum many-body systems, the ground states could be expressed in terms of a matrix product state (MPS), which can be expressed as 
\begin{align}
| \psi \rangle = \sum_{s_1,s_2,...,s_n} \textrm{tTr} \big[ 
A^{s_1} A^{s_2} \cdots A^{s_n} \big] | s_1  s_2 \cdots s_n \rangle
\end{align} 
where $s_k = 1,2,\cdots d_s$ for $k=1,2,\cdots n$ with $n$ the number of sites of the spin chain, and $A^{s_i}$ is the on-site tensor. 
For example,  $A^{s_k}_{l,r}$  are  rank-three tensor for 1D MPS as shown in Fig.~\ref{fig:rho_TN} (a). Moreover, we can also generalize this idea to represent quantum many-body operators, called matrix product operators (MPO), as shown in Fig.~\ref{fig:rho_TN} (b). We call $d_s$  the physical dimension and $\chi_{cut}$  the bond dimension. 
$\textrm{tTr}$ is to sum over all indices of tensors. Depending on the situation, we should tune $\chi_{cut}$ to capture the ground state's essential feature fully. 
Despite that, the $\chi_{cut}$ generally required for a good approximation of ground states still yields a more efficient representation of the quantum ground states than the typical dimensions of $d_s^N$. The validity of MPS ansatz is due to the area-law nature of quantum entanglement entropy of ground states and shall not hold for highly excited or high-temperature states. 

The MPS effectively expresses the ground states of quantum spin-chain models so that numerical methods can construct them more efficiently.  
We will use the time-evolving block decimation TEBD method  \cite{iTEBD2008} as described below to implement the imaginary time evolution operator to obtain the ground states of the two considered spin-chain models and then use them to evaluate the generating (characteristic) function for the work statistics.  
Furthermore, when checking the fluctuation theorem, one must prepare the initial states as Gibbs states, which are beyond primitive MPS. 

\subsection{Time-evolving block decimation}
The time-evolving block decimation (TEBD) provides an efficient way to simulate time evolution. 
We use the TEBD method to solve for the above MPS numerically, namely, by acting the imaginary time evolution operator $e^{-\tau H}$ on an initial state $| \psi_0 \rangle$ to determine the ground state and by acting the real-time evolution operator $e^{- i H t}$ to study the dynamics of the quantum lattice systems.

Denote the time-evolving state as $|\psi_t  \rangle = e^{-i t H}  |\psi \rangle$ for an initial state $|\psi \rangle$ with a given Hamiltonian $H$. For simplicity, we consider only the Hamiltonians made of arbitrary single-site and two-site terms, i.e., with nearest-neighbor interactions, so that they can be decomposed as $H = \sum_{k=1}^L h_{k,k+1} = H_{even} + H_{odd}$, where
$
 H_{even} \equiv \sum_{even \; k}^L h_{k,k+1}\;, \; 
 H_{odd} \equiv \sum_{odd \; k}^L h_{k,k+1}
$
and hence the commutator $\big[  H_{even},  H_{odd} \big] \neq 0$.
We then break the evolution operations $e^{-i t H}$ into a sequence of local gates using a Suzuki-Trotter expansion. 
For small enough $ \delta $,   the Suzuki-Trotter expansion of order  for $ e^{-i t H} $  can be written as
\begin{align}
\label{TEBD}
 e^{-i t H}& =  \big[ e^{- i\delta H} \big] ^{t/\delta} 
  =  \big[ e^{-i \delta ( H_{even} + H_{odd} )} \big] ^{t/\delta}  \notag\\
&  \approx  \big [  f_p ( e^{-i \delta  H_{even} },   e^{-\delta  H_{odd} } ) \big ] ^{t/ \delta} \;, 
\end{align} 
where $f_1=(x,y)= xy$,  $f_2=(x,y)= x^{1/2}yx^{1/2}$
for first and second-order expansions, respectively.
If we expand to higher-order terms, the Trotter error will decrease. The evolution operators can be expressed as a product of two-body gates using the Suzuki-Trotter expansion.
The simulation of evolution is achieved by updating the MPS by a sequential of alternating gate operations, i.e., alternating between $e^{-i \delta  H_{even} }$ and $e^{-i \delta  H_{odd} }$, as specified by Eq.~\eq{TEBD}. 
In this paper, we will implement this MPS-based TEBD method to evaluate the work characteristic function $G(u)$ for extracting average work done $\overline{W}$ and also evaluate the partition functions for extracting the free energies that appear on the right-handed-side of the fluctuation theorem.

\subsection{ Thermal density matrix as MPO}
As mentioned, the MPS is suitable for representing the ground states but not the highly excited states, such as high-temperature thermal states. This paper will mainly deal with the work statistics for ground states. On some occasions, we will also deal with thermal states, e.g., by checking the fluctuation theorem or the passivity of thermal states. A way to construct the thermal states based on MPS is the so-called algorithm of minimally entangled typical thermal states (METTS) \cite{stoudenmire2010minimally}, which adopts a Markov chain of product states to construct the thermal ensemble constrained by detailed balancing relation.

Instead of adopting METTS, a more direct way of constructing a thermal state based on MPS is to represent the density operator $e^{-\beta H}$ by an MPO. The partition function for evaluating free energy is to take a trace of this MPO. To construct such an MPO, one first prepares an initial identity MPO of bond dimension $\chi_{cut}=1$ to make such an MPO. The MPO for the thermal state is then obtained by acting with the operator $e^{-\beta H}$, which can be decomposed by TEBD and Suzuki-Trotter expansion into a sequence of a product of two-body gates, as shown in Fig. \ref{fig:rho_TN} (c).  The partition function can be obtained by tracing out the two physical indices of the MPO $e^{-\beta H}$ as shown in Fig.~\ref{fig:rho_TN}(d).
In Appendix \ref{app_a1}, we show the agreement on the work statistics obtained from METTS and MPO representation of thermal states. In the rest of the paper, we will adopt MPO to represent thermal states to proceed with the numerical calculations.

\begin{figure}
 \includegraphics[width=0.50\textwidth]{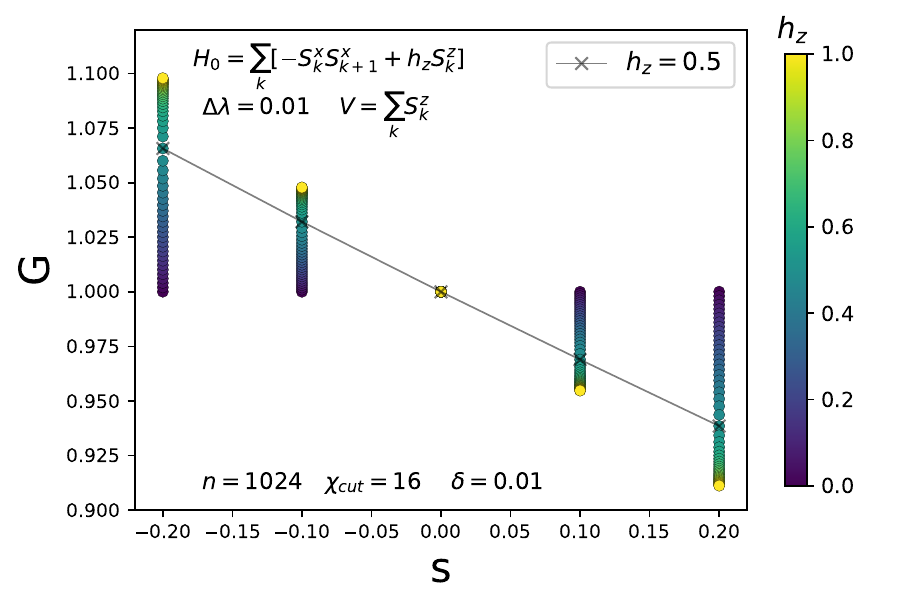}
 \caption{\small   
Work characteristic function $G(-is)$ under a sudden quench process implemented by Hamiltonian $H = H_0 +  \Delta \lambda\; V $ with $\Delta \lambda=0.01$  with $H_0$ of a $n=1024$ quantum Ising chain, i.e., \eq{q_XY} of $\gamma=1$, $h_x=0$ and $h_z \in [0,2]$, and $V=\sum_k S_k^z$.
 The results are obtained by the MPS-based TEBD method via the 1-st order Trotter expansions for $\delta=0.01$ with bond dimension $\chi_{cut}=16$. We show the results only for five values of $s$ but can connect them into a line by Lagrange interpolation. Each color represents one particular choice of $h_z$, with their values indicated by the color sidebar.
 The Lagrange interpolating lines at critical points are shown explicitly for $h_z=0.5$. 
 The slope of an interpolating line corresponds to the average work done $\overline{W}$, which is mostly negative. 
  }
 \label{fig:ZeroT_Gs_XY_XXZ}
\end{figure} 

\section{Numerical results for characterizing the quantum phase transitions}\label{sec:4}

In this section, we will present our numerical results to demonstrate that the average work done by the sudden quench can be the order parameter for the quantum phase transitions.  
The quantum systems we consider include quantum spin-$1/2$  and spin-$1$ chains. The quantum phase transition considered for the spin-$1/2$ chain is the Landau-Ginzburg type due to spontaneous symmetry breaking. On the other hand, the ones for the spin-$1$ chain model are the topological phase transitions, which can only be characterized by non-local order parameters. 

The average work done by sudden quench is given succinctly by Eq.~\eq{w_sq} or extracted from the characteristic function provided by Eq.~\eq{G_sq}.  In either case, we can implement the TEBD method based on MPS to evaluate it. Below, we will show the numerical results case by case.

\subsection{Quantum spin-$1/2$ chain in magnetic field}

\begin{figure*}
 \includegraphics[width=1.0\textwidth]{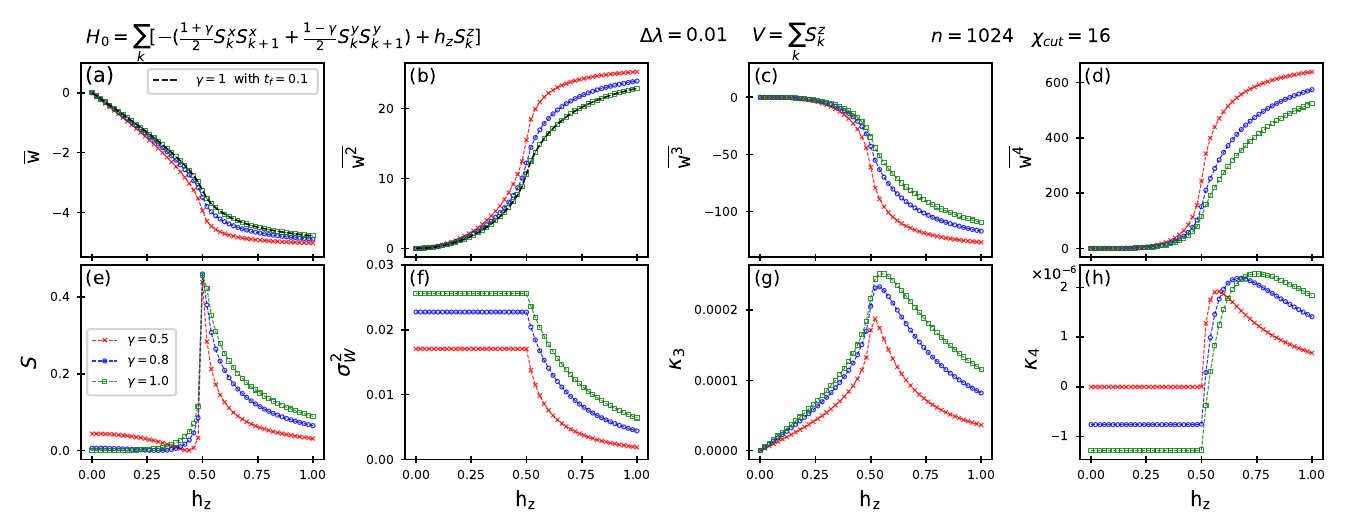}
 \caption{\small  
Moments and cumulants of work done by sudden quench with $V=\sum_k S_k^z$ and $\Delta \lambda=0.01$ on quantum Heisenberg XY Ising chain of size $n=1024$ as a function of the transverse field $h_z$ at zero temperature for $\gamma=0.5$ (red crosses), $0.8$ (blue circles) and $1.0$ (green squares). 
For comparison, the entanglement entropy $S$ is also shown. They are listed as follows: (a)  $\overline{W}$, (b)  $\overline{W^2}$, (c)  $\overline{W^3}$, (d) $\overline{W^4}$, (e) $S$, (f) $\sigma^2_W$, (g) $\kappa_3$ and (h) $\kappa^4$. The plots are obtained by the TEBD method with the parameters {$\chi_{cut}=16$ and $\delta=0.01$}, and $\overline{W^n}$ are obtained using Eq.~\eq{w_sq} by the MPO method. As expected, the higher moments or cumulants yield sharper change near the critical point so that the work statistics can indicate the quantum phase transitions in the Ising-like spin chains. To go beyond sudden quench and for comparison, we also extract $\overline{W}$ and $\overline{W^2}$ (black dashed lines) from the characteristic function Eq.~\ref{work_realtime} for the non-sudden quench process with $t_f =0.1$.
}  
 \label{fig:XY_work}
\end{figure*} 

We first consider the anisotropic Heisenberg XY spin-$1/2$ chain in a transverse magnetic field $h_z$ and a longitudinal field $h_x$. Its Hamiltonian is given by 
\begin{align}\label{q_XY}
H=&  \sum_{k=1}^n  \Big[ -  \big( { 1+\gamma  \over 2} S_k^x S_{k+1}^x + { 1-\gamma \over 2} S_k^y S_{k+1}^y \big) + h_x S_k^x +h_z S_k^z \Big]  
\end{align}
where $S_k^{x,y,z}$ are the site spin-$1/2$ operators, $k$ labels the number of sites. 
When $\gamma=1$, it reduces to the quantum Ising chain, and we will focus on the parameter range: $0\le \gamma \le 1$ and $h_x=0$. 
The ground state of this model can be exactly solved and exhibits three phases of the Landau-Ginzburg-Wilson paradigm: the oscillatory phase $(O)$, the ferromagnetic phase $(F)$, and the paramagnetic phase $(P)$~\cite{henkel2013conformal}.
The ground state stays in the $O$ phase for small $h_z$ until $h_z={1\over 2}\sqrt{1-\gamma^2}$ and then changes to the $F$ phase as a crossover transition. Further increasing $h_z$ up to $h_z=0.5$, the ground state will change to $P$ phase, which is a second-order quantum phase transition in the thermodynamic limit with $\langle S^x \rangle$ as the order parameter, i.e., $\langle S^x \rangle$ is nonzero in $F$ (or $O$ if $\gamma=0$) phase but is zero in $P$ phase. 
Below, we will consider the phase transition between the ferromagnetic and paramagnetic phases.

Here, we apply the MPS-based TEBD method to calculate the characteristic function $G(u)$ of Eq.~\eq{work_realtime} for this spin chain model with $\gamma=1$ and $h_x=0$ at zero temperature but varying the transverse field $h_z$. We aim to check if the work done $\overline{W}$ can be used as an order parameter for the quantum phase transition, i.e., from $F$ (or $O$ if $\gamma=0$) phase to $P$ phase.  
For simplicity, we consider the sudden quench process implemented by the Hamiltonian of Eq.~\eq{H(t)} and Eq.~\eq{couple_sq} with a chosen jump of the coupling constant,  i.e., $\Delta \lambda=0.01$.
The results are shown in Fig. \ref{fig:ZeroT_Gs_XY_XXZ} and \ref{fig:XY_work}.

The Fig.~\ref{fig:ZeroT_Gs_XY_XXZ} shows the characteristic function $G(-is)$ of a $n=1024$ Ising chain for $s \in [-0.2,0.2]$ with $\gamma=1$ and $h_x=0$ at various values of $h_z$ in the  initial Hamiltonian $H_0$. 
For simplicity, we pick five values of $s$ but vary $h_z$ continuously with its value indicated by the color bar attached aside. In particular, the five points belonging to $h_z=0.5$ are denoted by the crosses and joined into a line by Lagrange interpolation.  We see that the slope of this line is negative, which implies that the average work done $\overline{W}=\lim_{s\rightarrow 0} {\partial G(-is) \over \partial s}$ is negative. For other values of $h_z$ in the initial Hamiltonian $H_0$, we can see that the slope of the line connected by the five points of the same $h_z$ starts with a positive value for $h_z=0$, then gradually turns negative as $h_z$ increases and finally converges to a constant negative value. 

The above results and further related calculations can be translated into phase diagrams as shown in Fig.~\ref{fig:XY_work}   for the quantum phase transition from the $F$ (or $O$) phase to the $P$ phase.  
These phase diagrams are characterized by the entanglement entropy $S$, the average work done, $\overline{W}$,  and its associated higher moments and cumulants, under a sudden quench process with $V=\sum_k S_k^z$ and $\Delta \lambda=0.01$. We see that $S$ has a sharp peak near $h_z=0.5$, which implies a stronger correlation near the critical point at $h_z=0.5$. Similarly, the work statistics denoted by $\overline{W}$ and the higher moments and cumulants also show the shaper changes near $h_z=0.5$. As expected, the higher moments and cumulants show sharper changes at the critical point. This implies that the work statistics can indicate the phase transitions. Interestingly, the cumulants of even orders show a level-off for the $F$ (or $O$) phase with $h_z\le 0.5$, and similar behavior for $\sigma^2_W$ has been observed earlier obtained by the exact solution of Ising chains \cite{bayocboc2015exact}. This constancy could be associated with some underlying symmetry. However, the quantities $\langle V^n \rangle$ with $V=\sum_k S_k^z$, related to the cumulants of work, are different from the order parameter $\langle \sum_k S_k^x \rangle$ for the transition from $F$ (or $O$) phase to the $P$ phase.

In addition, in Appendix~\ref{app_a3}, we give a consistency check to show  in Fig. \ref{fig:Ising_consistent_test}  that the above numerical results of $\overline{W}$ and $\overline{W^2}$ obtained  from either Eq.~\eq{w_avg} or Eq.~\eq{w_sq} agree.

 \begin{figure*}
 \includegraphics[width=1.00\textwidth]{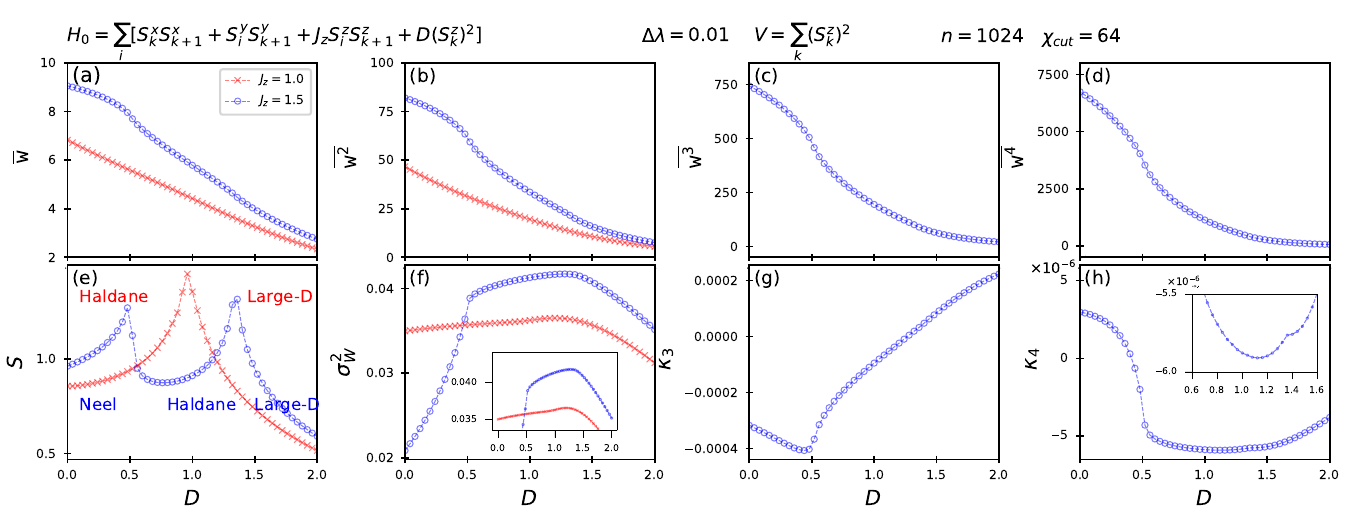}
 \caption{\small 
Moments and cumulants of work done by sudden quench with $V=\sum_k (S_k^z)^2$ and $\Delta \lambda=0.01$ on quantum Heisenberg XXZ Haldane chain of size $n=1024$ as a function of the coupling parameter $D$ at zero temperature for $J_Z=1.0$ (red crosses) and $1.5$ (blue circles). For comparison, the entanglement entropy $S$ is also shown. The listing order for $\overline{W^{m=1,\cdots, 4}}$, $\sigma^2_W$,  $\kappa_{\ell=3,4}$ and $S$, the subfigure captions and the numerical implementing methods are the same as the one in  Fig. \ref{fig:XY_work}. The quantum critical points are indicated by the sharper behavior of $S$, which are at (i) $(J_z, D)=(1.5,0.48)$, (ii) $(J_z, D)=(1.5, 1.34)$ and (iii) {$(J_z, D)=(1.0,1.0)$}. Note that (i) is the transition point from the ordered phase to the topological phase, but (ii) and (iii) are the purely topological ones without any order parameter and should be more difficult to detect. For (i), $\overline{W^{m=1,\cdots, 4}}$ show very mild behavior, but cumulants show shaper changes as the order of cumulants goes higher. However, for (ii) and (iii), only $\sigma^2_W$ and $\kappa_{4}$ show mild crossover (see the enlargement insets in the subfigures (f) and (h)), and all the other moments/cumulants in this figure show no obvious changes. Overall, we can conclude that the work statistics can barely capture the purely topological phase transitions.  {The plots are obtained by the TEBD method with the parameters $\chi_{cut}=64$ and $\delta=0.01$.} }
 \label{fig:XXZ_Work}
\end{figure*} 

\subsection{Anisotropic quantum spin-$1$ chain}

To show that our proposal can also work for the quantum phase transitions of the non-Landau-Ginzburg-Wilson type,  we consider a particular model of such type, the anisotropic quantum XXZ spin-$1$ chain described by the following Hamiltonian:
\begin{align}\label{Haldane_1}
 H=  \sum_{k=1}^n \Big[  S^x_k  S^x_{k+1} + S^y_k  S^y_{k+1} + J_z  S^z_k  S^z_{k+1}  
 + D  ( S^z_{k})^2 \Big],
\end{align}
where $S_k^{x,y,z}$ are the site spin-$1$ operators, and the parameter $D$ denotes the uni-axial anisotropy. When $J_z=1$, it reduces to the so-called Haldane chain. 
The ground-state phase diagram of this model consists of six different phases~\cite{Hatsugai_XXZ_1991, Chen_XXZ_2003}. 
Here, we focus on phase transitions among three phases: the large-$D$ phase, the N'eel phase, and the Haldane phase characterized by nonzero string-order parameters.
At large $D$, the model is in a trivial insulator phase. Namely, the ideal large-D phase is $| 000\cdots \rangle $.  On the other hand, the ideal N'eel phase is $| 1,-1,1,-1 \cdots \rangle$ or $| -1,1,-1,1 \cdots \rangle$ with spontaneous nonzero expectation values of the staggered magnetization. 

The Haldane phase is one of the symmetry-protected topological (SPT) phases. The ground states of such phases have nontrivial patterns of quantum entanglement. They cannot continuously connect to trivial product states without closing the gap or breaking the protecting symmetry. Thus, the SPT phases preserve the global symmetry of the Hamiltonian so that some spontaneous symmetry-breaking local order parameters cannot characterize the transition from the SPT-nontrivial phase to the SPT-trivial one. This is in contrast to the cases of the Landau-Ginzburg type. Instead, the topological phases,  such as the Haldane phase, are characterized by fractionalized edge excitations, which some nonlocal order parameters, such as the expectation value of some string-like operator, can measure.

As before, we can characterize the quantum phase transitions of this model by entanglement entropy $S$, the average work done $\overline{W}$ and its higher moments/cumulants by a sudden quench with $V = \sum_k (S_k^z) ^2$ and $\Delta \lambda=0.01$, and by the associated higher moments and cumulants. The results are shown in Fig.~\ref{fig:XXZ_Work}, which are again obtained numerically using the MPS-based TEBD method.  We see that the entanglement entropy $S$ shows the peaks around three quantum critical points, which are at (i) $(J_z, D)=(1.5,0.48)$, (ii) $(J_z, D)=(1.5, 1.34)$ and (iii) {$(J_z, D)=(1.0, 1.0)$}. The critical point (i) is the transition point from the ordered Neel phase to the topological Haldane phase and is partly with local order parameters. On the other hand, the critical points (ii) and (iii) are from the Haldane phase to the large $D$ phase, and no local parameter exists to characterize it. Therefore, these transitions are purely topological types and should be more difficult to detect. Indeed, this is what we see from Fig.~\ref{fig:XXZ_Work}. For (i), $\overline{W^{m=1,\cdots, 4}}$ show very mild behavior, but cumulants show shaper changes as the order of cumulants goes higher. However, for (ii) and (iii), only $\sigma^2_W$ shows mild crossover, and all the other moments/cumulants in this figure show no obvious changes. It is strange why the higher cumulants do not show sharper behavior for critical points (ii) and (iii). This could be due to the need for a higher bond dimension to capture the behavior of higher cumulants. Overall, we can conclude that the work statistics can barely capture the purely topological phase transitions. Since work statistics involve all multipoint correlations, and could be used as some non-local order parameter. Thus, we expect the higher moments and cumulants of the work statistics to display shaper behavior around the topological phase transitions,  although it is quite computationally costly to obtain higher moments and cumulants due to the need for large bond dimensions. 

Also, in Appendix~\ref{app_a3}, we give a consistency check to show  in Fig. \ref{fig:XXZ_consistent_test}  that the above numerical results of $\overline{W}$ obtained  from either Eq.~\eq{w_avg} or Eq.~\eq{w_sq} agree. 
However, the  $\overline{W^2}$ obtained  from either Eq.~\eq{w_avg} or Eq.~\eq{w_sq} can't match well. The larger bound dimension $\chi_{cut}$ might improve the physical properties near critical points.

\section{Benchmarking the real-time evolving methods by Jarzynski's equality}
\label{sec:5}

\begin{figure}
 \includegraphics[width=0.48\textwidth]{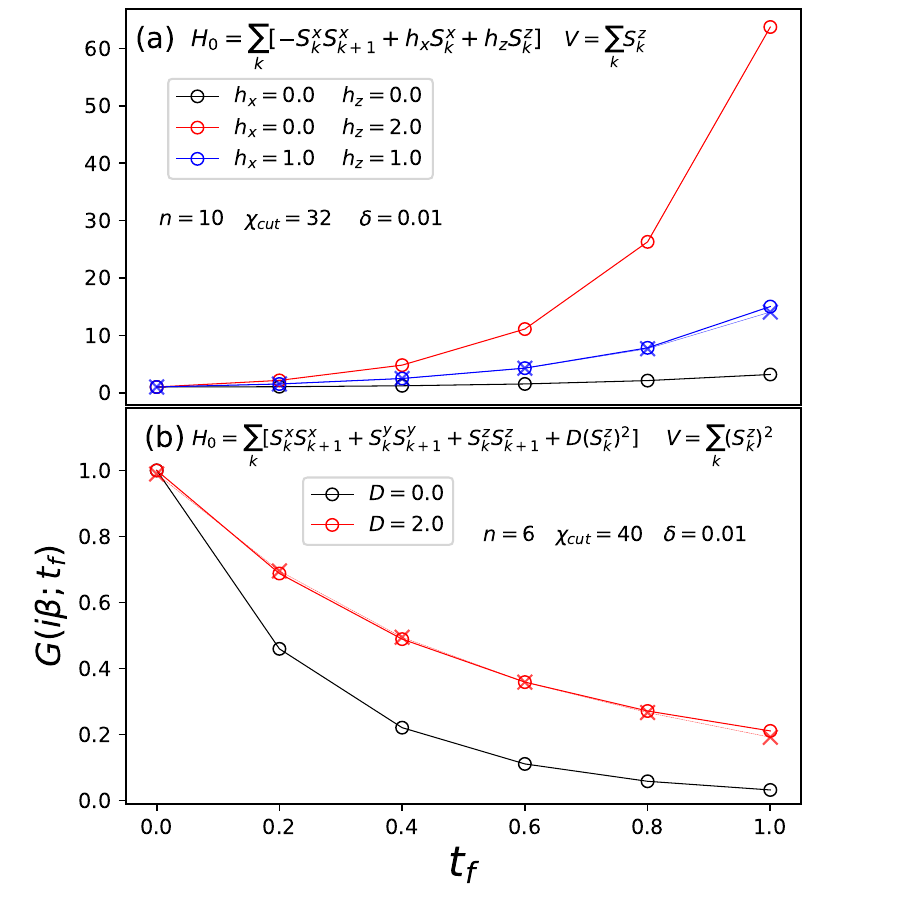}
 \caption{\small  
Work characteristic function $G(i\beta; t_f)\vert_{\beta=1}$ implemented by Hamiltonian $ H = H_0 + \lambda(t) V$ with $\lambda(t)=t$, for 
  (a) $n=10$ Ising chain, and (b) $n=6$ Haldane chain, with their $H_0$ and $V$ shown on the corresponding subfigures. 
We obtain the results by using the ED method (circles) and the MPS-based TEBD method (crosses) with the parameters $\chi_{cut}=40$ and $\delta =0.01$. In some cases, both results from ED and MPS agree. Note that $G(i\beta; t_f)\vert_{\beta=1}$  increases with time for (a) but decreases for (b). 
}
\label{fig:Gu_check}
\end{figure} 

We now adopt the Jarzynski equality of Eq.~\eq{Jarzynski_1}  dictated by the real-time correlator expression of the work characteristic function $G(i\beta;t_f)\vert_{\beta=1}$, i.e., Eq.~\eq{work_realtime}, to benchmark the numerical accuracy of the real-time evolution of two quantum spin chain models considered above. The numerical error mainly comes from evaluating $G(i\beta;t_f)$ and accumulates as $t_f$ grows. For the small-size cases, we can use either the Exact Diagonalization (ED) method or the MPS-based TEBD method. 
For large-size cases,  we can only use the TEBD method.  As before, the non-equilibrium process is implemented by the Hamiltonian of Eq.~\eq{H(t)}, and here, we adopt the linear profile of coupling constant, i.e., $\lambda(t)=t$. The results for the small-size cases are shown in 
Fig.~\ref{fig:Gu_check} for (a) $n=10$ Ising chain of fixed $h_{x,z}$ and with $V = \sum_i S_i^z$ and (b) $n=6$ Haldane chain of fixed $D=2$ and with  $V = \sum_i (S_i^z)^2$.   Fig.~\ref{fig:Gu_check} shows that as $t_f$ increases, the curve of $G(i\beta;t_f)\vert_{\beta=1}$ of the Ising chain increases; on the other hand, the curve in the Haldane chain decreases.

\begin{figure}
\includegraphics[width=0.48\textwidth]{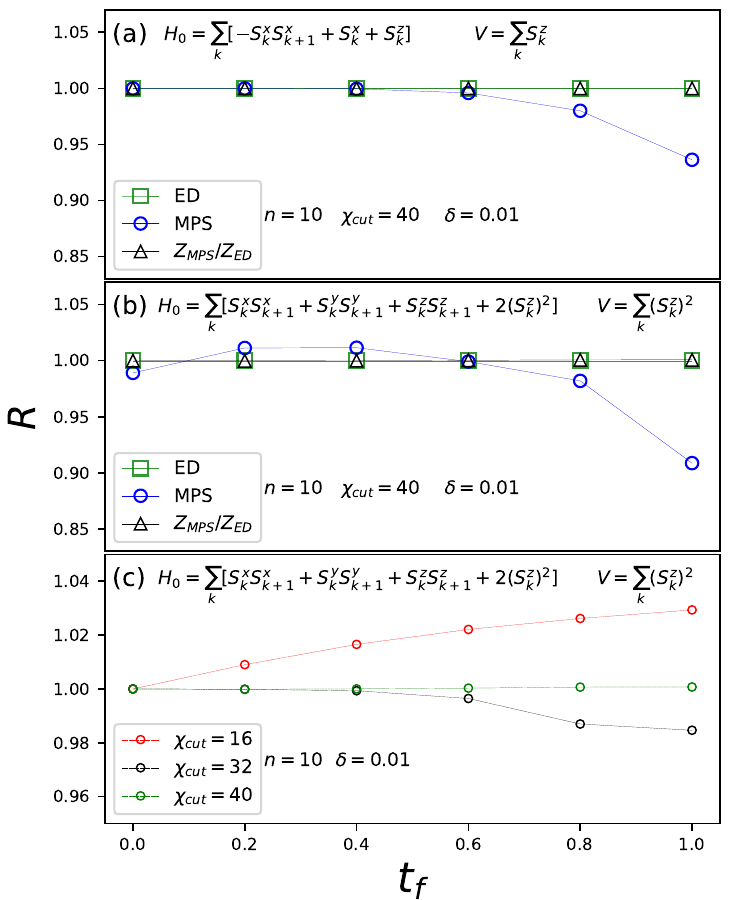}
 \caption{\small 
 Benchmarking the ED and TEBD methods by the ratio $R(t_f)={G(i\beta;t_f) \over Z(t_f)/Z(0)}\Big \vert_{\beta=1}$ from Jarzynski equality for subfigure (a) $n=10$ Ising chain with $h_x=h_z=1$ and $V=\sum_k S_k^z$; and 
 subfigure (b) $n=6$ Haldane chain with $D=2$ and $V=\sum_k (S_k^z)^2$. In subfigure (c), we show the dependence of the resultant $R(t_f)$ of $n=6$ Haldane chain on the bond dimension $\chi_{cut}$ of MPS. As expected, larger $\chi_{cut}$ yields better results as expected. The results from ED are indicated by squares, and from TEBD, they are indicated by circles. Besides, we also show the ratio ${Z(t_f)\over Z(0)}|_{\rm MPS} \Big/ {Z(t_f)\over Z(0)}|_{\rm ED}$
 (triangles in subfigures (a) and (b)) to demonstrate the accuracy of TEBD in evaluating the partition functions. In all the above, the coupling constant has the linear profile, i.e., $\lambda(t)=t$.
 If there is no numerical error, $R=1$, and the deviation benchmarks the real-time numerical errors. 
 For ED, we see that $R$ remains equal to one. On the other hand, for MPS, we see that the real-time numerical errors start to escalate around $t_f=0.7$. However, the ratio of partition functions remains accurate. 
 }
 \label{fig:JK_check}
\end{figure} 

\begin{figure}
\includegraphics[width=0.48\textwidth]{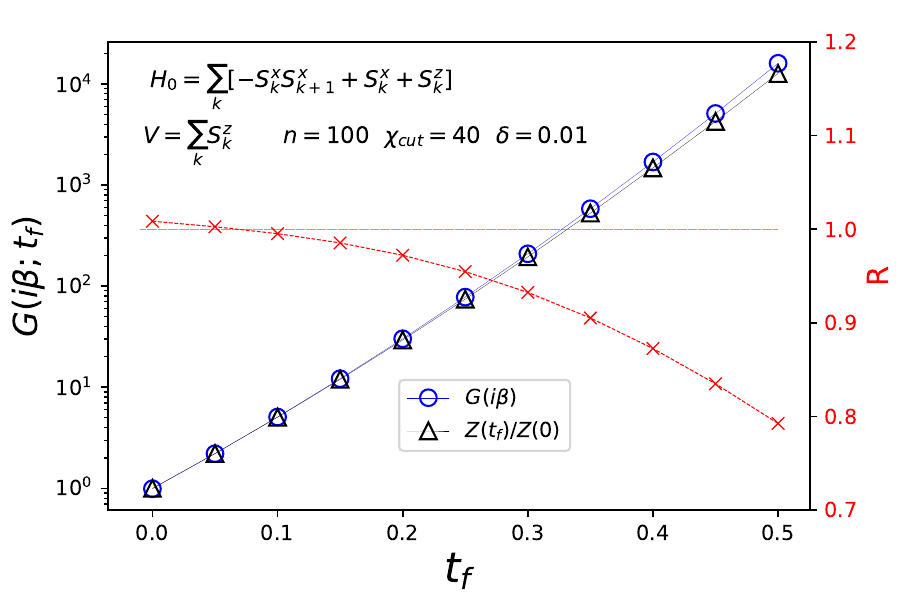}
 \caption{\small 
 Benchmarking the numerical accuracy with Jarzynski equality for $n=100$ Ising chain with the same $H_0$, $V$, $\chi_{cut}$ and $\delta$ as in Fig. \ref{fig:JK_check}(a). The results for $G(i\beta;t_f)\vert_{\beta=1}$, $R(t_f)$ and  ${Z(t_f) \over Z(0)}\Big \vert_{\beta=1}$ are denoted by blue circles, red crosses, and black triangles, respectively.  
 Due to the large size, ED is unavailable, and the results are obtained by the TEBD method. We see that the real-time numerical errors start to escalate around $t_f=0.13$, which is far shorter than the $n=10$ case as expected. However, the ratio of partition functions remains accurate as in the $n=10$ case. 
 }
 \label{fig:JK_ising_L100}
\end{figure} 

Next, we use Jarzynski equality to characterize the numerical error of real-time evolution and benchmark the corresponding numerical methods. This can be quantified by the ratio $R(t_f)={G(i\beta;t_f) \over Z(t_f)/Z(0)}\vert_{\beta=1}$ defined in Eq.~\eq{benchmark_R}, which is the ratio between either side of Jarzynski equality. If there is no numerical error, the Jarzynski equality is obeyed so that $R=1$. Otherwise, the $|1-R|$ size can characterize the numerical error.

We can evaluate  the partition function ratio ${Z(t_f) \over Z(0)}\Big \vert_{\beta=1}$ by ED and TEBD methods. As expected, both methods can be accurate for such non-dynamical quantity. Combined with the previous results for $G(i\beta;t_f)\vert_{\beta=1}$, we can evaluate $R$. The results for small-size quantum chains are shown in Fig.~\ref{fig:JK_check} with subfigure (a) for $n=10$ quantum Ising chain and subfigure (b) for $n=6$ Haldane chain.
We see that $R$ remains one for the ED method. 
However, for the MPS-based method, the numerical errors escalate around some critical moment, i.e., $(t_f)_c=0.7$ as shown in Fig.~\ref{fig:JK_check} (a) and (b). 
Moreover, we also evaluate and show the ratio of the ratios of the partition functions obtained by ED and MPS-based methods, i.e., ${Z(t_f)\over Z(0)}|_{\rm MPS} \Big{/} {Z(t_f)\over Z(0)}|_{\rm ED}$ (black triangles in Fig.~\ref{fig:JK_check} (a) and (b)). In Fig.~\ref{fig:JK_check} (c), we show the dependence of the resultant $R(t_f)$ of $n=6$ Haldane chain on MPS's bond dimension $\chi_{cut}$. The larger $\chi_{cut}$ yields better results as expected. In this case, $\chi_{cut}\ge 40$ is needed to arrive at an accurate $R=1$ result for $t_f\le 1$.

When the quantum spin chain size becomes large, the ED method is unavailable, and we can only use the MPS-based method to evaluate $R$. 
For comparison, we evaluate the benchmarking factor $R$ for $n=100$ Ising chain with the same $H_0$, $V$, $\chi_{cut}$ and $\delta$ as in the $n=10$ case of Fig. \ref{fig:JK_check}(a).
The results of $G(i\beta;t_f)\vert_{\beta=1}$, $R(t_f)$ and  ${Z(t_f) \over Z(0)}\Big \vert_{\beta=1}$ are shown in Fig. \ref{fig:JK_ising_L100}. As we can see, $G(i\beta;t_f)\vert_{\beta=1}$ grows exponentially with $t_f$, which should cause large numerical errors. Indeed, the real-time numerical errors characterized by $|1-R|$ start to escalate around the critical moment $(t_f)_c=0.13$. This is far shorter than $(t_f)_c=0.7$ of the $n=10$ case, as we expect that the numerical errors accumulate more quickly for systems of larger sizes.

\section{Numerical results for examining the passivity of thermal/ground states}\label{passivity_sec}

\begin{figure}
\includegraphics[width=0.48\textwidth]{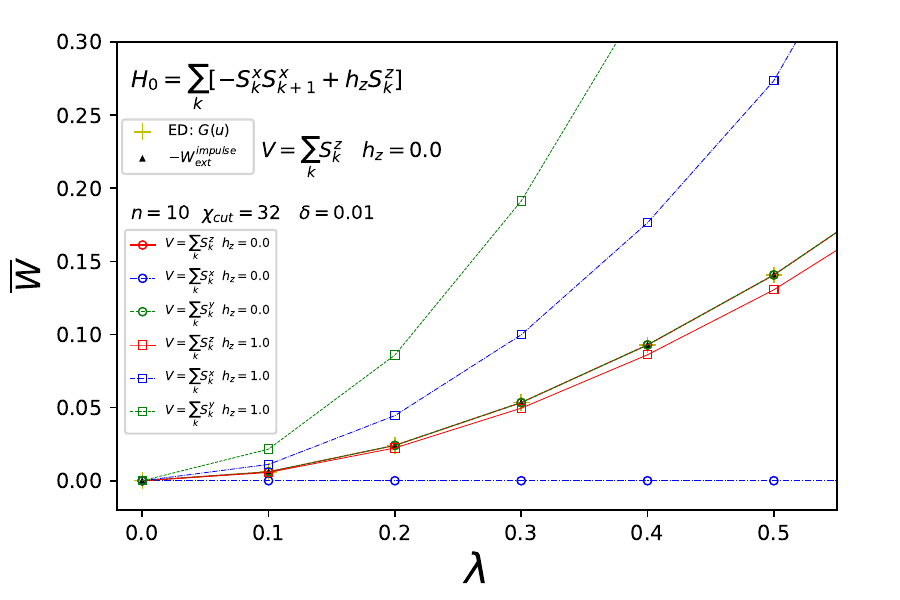}
 \caption{\small 
 Demonstration of passivity of thermal states in $n=10$ quantum Ising chain under impulse process implemented by the Hamiltonian of Eq.~\eq{H(t)} and Eq.~\eq{couple_impulse} with $H_0$ of Eq.~\eq{q_XY} with $h_x=h_z=0$ and with $V=\sum_k S_k^x$, $\sum_k S_k^y$ or $\sum_k S_k^z$. The resultant average work done per site $\overline{w}^{\rm impulse}$ for $\beta=1$ are obtained by ED and TEBD. Its positiveness for all considered  $V$'s and $\lambda \in [0,0.5]$ demonstrates the passivity of thermal states as pointed out in Eq.~\eq{2ndLaw_passive}. 
 }
\label{fig:passivity_cIaing}
\end{figure} 

\begin{figure}[]
 \includegraphics[width=0.45\textwidth]{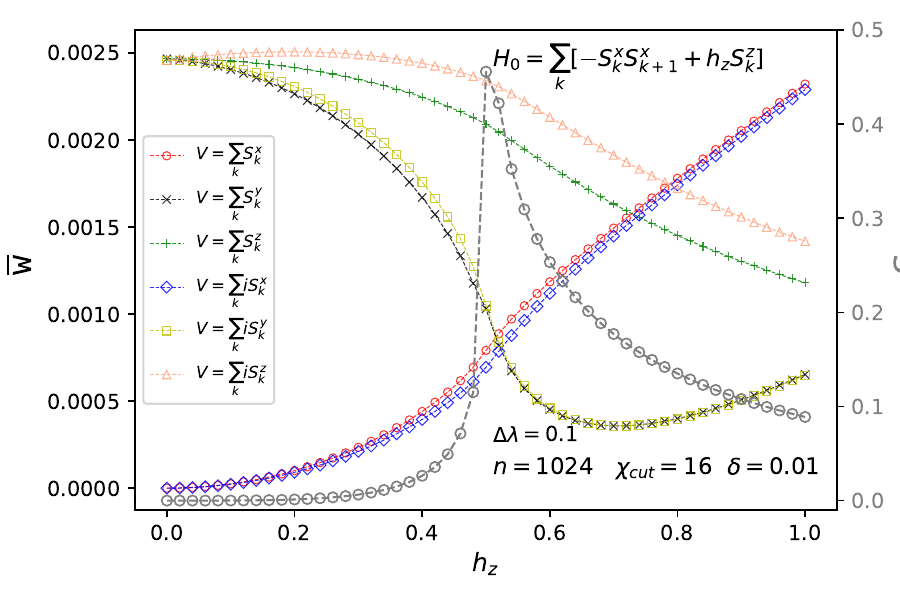}
 \caption{\small 
  Plot of $\overline{w}^{\rm impulse}$ to show the passivity of ground states of $n=1024$ quantum Ising chain with $h_x=0$ but varying $h_z \in [0,1]$ under impulse process implemented by Hamiltonian of Eq.~\eq{H(t)} and Eq.~\eq{couple_impulse} with $\lambda=0.1$ and Hermitian actions $V=\sum_k S_k^x$ (red circles ), $\sum_k S_k^y$ (black cross), $\sum_k S_k^z$ (green cross), and non-Hermitian actions $V=\sum_k i S_k^x$ (blue diamonds), $\sum_k i S_k^y$ (yellow square), $\sum_k i S_k^z$ (pink triangles). Again, we see that $\overline{W}^{\rm impulse}$ can also indicate quantum phase transition by mild crossover behaviors when compared with the phase diagram of the entanglement entropy of initial state (gray circle).  }
\label{fig:passivity_QIsing_W}
\end{figure}

\begin{figure*}
\includegraphics[width=1.0 \textwidth]{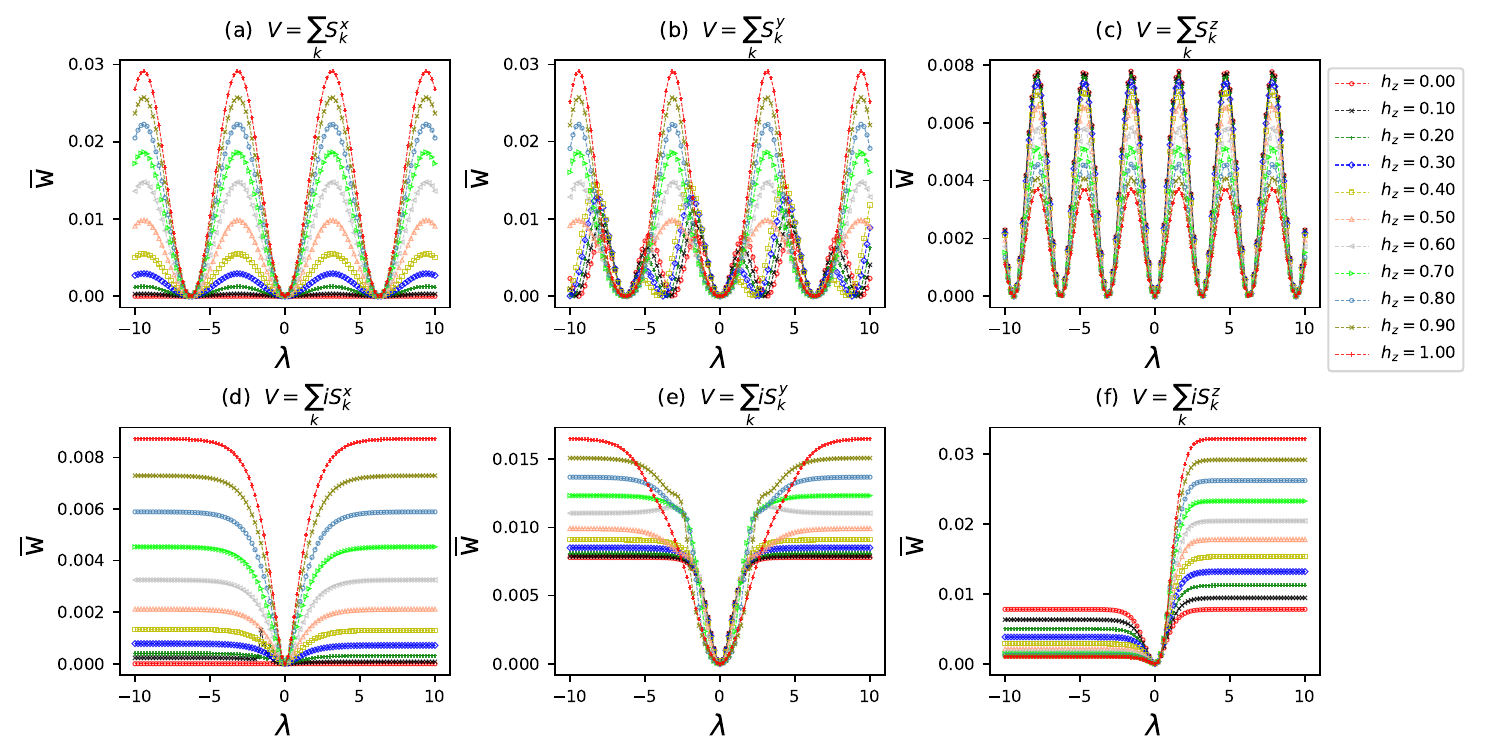}
 \caption{\small 
    Plot of $\overline{w}^{\rm impulse}$ to show the passivity of ground states and its pattern as a function of the coupling $\lambda$ of $n=1024$ quantum Ising chain.  We fix $h_x=0$ but vary $h_z \in [0,1]$ to obtain various ground states, and also vary $\lambda \in [-10,10]$ under impulse process implemented by Hamiltonian of Eq.~\eq{H(t)} and Eq.~\eq{couple_impulse} with the Hermitian actions (a) $V=\sum_k S_k^x$, (b) $V=\sum_k S_k^y$, (c) $ V= \sum_k S_k^z$ in the top row, and the non-Hermitian actions (d) $V= i \sum_k S_k^x$, (e) $V=i\sum_k S_k^y$ and (f) $ V= i\sum_k S_k^z$ in the bottom row. }
\label{fig:passivity_ising}
\end{figure*}

We now can consider the passivity of a given initial state undergoing a (cyclic) impulse process. Two natural candidate categories are thermal or ground states.  
The passivity of (relativistic) thermal states was shown to be ensured \cite{Pusz:1977hb,1978JSP....19..575L,goldstein2013second}. 
For the thermal states of the non-relativistic systems, such as the quantum spin chains, as argued in Eq.~\eq{2ndLaw_passive}, the passivity should be guaranteed by the fluctuation theorem, i.e., the second law of thermodynamics.  Here, we demonstrate this is the case for the $n=10$ Ising chain. 
The result is shown in Fig.~\ref{fig:passivity_cIaing}, from which we see that the average work done per site $\overline{w}^{\rm impluse}=-w^{\rm impulse}_{\rm ext}$ for the thermal state at $\beta=1$ is always positive for various $V$ and $\lambda$.  

We now turn to the issue of passivity for the ground states under the cyclic impulse processes. As shown in section \ref{sec:IID}, the passivity of ground states is guaranteed to be passive under the Hermitian action, i.e., $V^{\dagger}=V$ by the variational principle. This is because the average work extraction from a cyclic process can be understood as the energy difference between the ground state and the excited state driven by $V$. It is always negative as the variational principle guarantees. On the other hand, there is no such simple interpretation for the average work extraction if the action $V$ is non-Hermitian, i.e., one cannot apply the variational principle to check the passivity of ground states. For the latter cases, it is then interesting to examine the passivity by evaluating average work extraction numerically. 

As a startup, we numerically examine the passivity of ground states for various Hermitian or non-Hermitian actions but small $\lambda$, i.e., $\lambda=0.1$. In Fig. \ref{fig:passivity_QIsing_W}, We plot the average work done per site $\overline{w}^{\rm impluse}=-w^{\rm impulse}_{\rm ext}$ as the function of $h_z$, which labels the ground states of the Ising-like chain under various Hermitian and non-Hermitian actions. A similar result for the Haldane-like chains is also given in Fig. \ref{fig:passivity_QXXZ_W} of Appendix \ref{App:passive}.  To obtain these results, we first evaluate $G(-is)$ by the MPS-TEBD method, and then from it extract $\overline{w}^{\rm impluse}=-w^{\rm impulse}_{\rm ext}$ from $G(-is)$. The ground states considered are all passive under the various Hermitian and non-Hermitian actions described in Fig. \ref{fig:passivity_QIsing_W}. This is a consistency check for our numerical method with the discussion in section \ref{sec:IID} for examining the passivity of ground states of Ising-like chains.

By exploiting the numerical method's power in checking the ground states' passivity, we now consider the cases with quite an extensive range of $\lambda \in [-10,10]$ for both Hermitian and non-Hermitian actions. The results for the Ising-like chains are shown in Fig. \ref{fig:passivity_ising}, and the similar results for the Haldane-like chains and the more general non-Hermitian actions in the Ising-like chains are shown in Fig. \ref{fig:passivity_XXZ_all} and Fig. \ref{fig:passivity_ising_v2} of Appendix \ref{App:passive}, respectively. Since the ones shown in the Appendix \ref{App:passive} bear similar features as in Fig. \ref{fig:passivity_ising}, we will not discuss them in the main text. The top row of Fig. \ref{fig:passivity_ising} shows the average work done per site on the ground states labeled by the values of $h_z \in [0,1]$ for the Hermitian actions $V=\sum_k S_k^{x,y,z}$, and the bottom row shows the results for the non-Hermitian actions $V=i \sum_k S_k^{x,y,z}$.  For the Hermitian cases, we see a quite interesting feature that $\overline{W}^{\rm impulse}$ is periodic with respect to the coupling $\lambda$ with the period equal to $2\pi$ or $\pi$. This periodic behavior has been discussed in section \ref{sec:IID} due to the peculiar feature of the impulse process. We will not see such behavior for the cyclic process with a finite time duration. For the non-Hermitian case, we first see that all the ground states considered are passive as in the Hermitian cases. However, $\overline{W}^{\rm impulse}$ is no longer periodic with respect to $\lambda$ nut levels off as $\lambda$ becomes large. In both cases, the average work done is bounded no matter how large the coupling $\lambda$ is. Again, this is due to the zero time duration of the impulse process, so the average work done is limited to such a tiny time duration. However, it is unclear why the passivity remains intact for non-Hermitian action, especially without the guarantee by the variational principle. This issue deserves future study to explore.

\section{Conclusions}\label{sec:6}

In this paper, we apply the numerical method for real-time evolution, such as Exact Diagonalization (ED) and MPS-based TEBD, to quantum spin lattice models to study the work statistics. With the power of MPS formalism, we can evaluate the chains up to $1024$ sites, which can effectively suppress the finite size effect.  We focus on three aspects: (i) study the behaviors of the moments and cumulants of the work statistics near the quantum phase transitions and examine their capability to indicate quantum critical points; Our results up to the fourth cumulant show that the work statistics can detect the quantum phase transitions charactered with local order parameters, but just barely for the topological phase transitions;  (ii) we propose to adopt Jarzynski equality as the benchmark for the accuracy of the numerical real-time evolution methods; (iii) we examine the passivity of thermal states and ground states of quantum spin chains under some cyclic impulse processes. Our numerical results show that all the ground states are passive under both Hermitian and non-Hermitian actions considered in this work. Although the variational principle ensures the passivity of ground states under Hermitian actions, it is not the case for non-Hermitian actions.  It is interesting to explore the reason for the passivity of ground states under non-Hermitian action seen in this paper, and also explore the possibility of active ground states by more general actions. Once the active ground states exist, we may adopt them to implement the quantum engine naturally to extract quantum work in the cyclic processes.

The quantum spin lattice with nearest neighbor interactions can be the natural system for performing quantum simulation and can serve as accurate tests for work statistics such as fluctuation theorem. Due to the statistical nature of quantum work, it has remained quite mysterious since its proposal decades ago. With our demonstration of numerical studies for the realistic many-body systems, one can explore more different perspectives of work statistics.  For example, when using pure states as the initial states for the work statistics, it may need more subtle treatment than the two-point measurements to preserve the quantum coherence and explore its role in the fluctuation theorem. We may hope to adopt the real-time evolution method, such as the MPS-based one, to investigate their role in some specific quantum tasks.

\begin{acknowledgments}
We thank Masahiro Hotta and Jhh-Jing (Arthur) Hong for their helpful discussions. FLL is supported by the Taiwan Ministry of Science and Technology through Grant No.~112-2112-M-003 -006 -MY3. CYH is supported by the Taiwan Ministry of Science and Technology through Grant No.~112-2112-M-029 -006. 
\end{acknowledgments}

\appendix


\section{ Tentative study of the density of state in  Eq.~\eq{W_direct}} \label{app_a2}

In section \ref{work done of sudden quench}, we argue why the average work done $\overline{W}$ can be used as the local order parameter for quantum phase transition.  As $\overline{W}$ can be expressed in \eq{W_direct} as the average over the eigen-energies of the final Hamiltonian, it will have a sudden change when crossing the quantum critical point because the probabilities of states $p(f)$ in \eq{W_direct} for the gapped and the gapless phases have pretty different behaviors at low energy of an extensive system. 

\begin{figure}
 \includegraphics[width=0.45\textwidth]{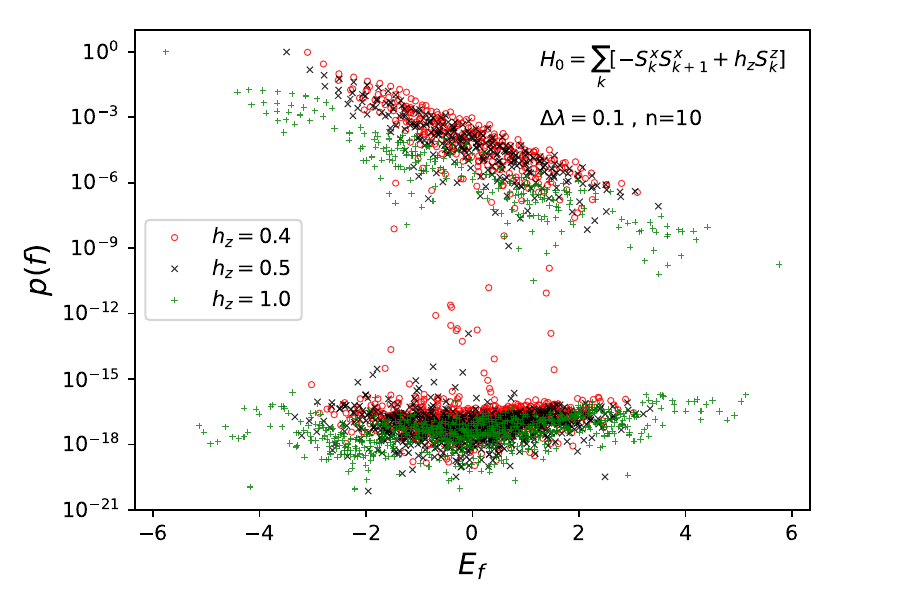}
 \caption{\small 
 The ED results of weight $p(f)$ from  Eq.~\eq{W_direct} are plotted as a function of the eigen-energy $E_f$ of the Hamitain $H_+ = H_0+ \Delta \lambda V$ for the n = 10 quantum Ising chains with $h_z =0.4$, $0.5$, $1.0$ under a sudden quench process with $V = \sum_k S_k^z $ and $\Delta \lambda = 0.1 $. We can see that the $p(f)$'s for $h_z=0.5$ (black crosses) and $h_z=0.4$ (red circles) have a more focused energy spectrum than the one for $h_z=1.0$ (green pluses). Note that for $h_z=0.5$ ($h_z=0.4$), the initial (final) Hamiltonian is in the critical phase because $\Delta \lambda=0.1$. Thus, the more focused energy spectrum for these two cases implies a tendency for the effect of ground-state degeneracy.
}
 \label{fig:pf_distribution}
\end{figure}

Therefore, it is interesting to calculate $p(f)$ and examine the behaviors of the gapped and gapless phases of an extensive spin chain. Unfortunately, the MPS-based method used for the main calculations of the long spin chain can not applied to the excited states. We can only study $p(f)$ tentatively by using ED method for the short spin chains. In this case, we will miss the ground state degeneracy for the gapless phases due to the sizeable finite-size effect. Despite that, we still present the ED results of $p(f)$ as a function of eigen-energy $E_f$ of the $n=10$ quantum spin chains with $h_z =0.4$, $0.5$, $1.0$ under a sudden quench process with $V = \sum_k S_k^z$ and $\Delta \lambda = 0.1$. The result is shown in Fig. \ref{fig:pf_distribution}. As expected, due to the large finite size effect for $n=10$, it is hard to see the effect of ground state degeneracy on $p(f)$ for either 
$h_z=0.5$ (black crosses) with the initial Hamiltonian in the critical phase or $h_z=0.4$ (red circles) with the final Hamiltonian in the critical phase as $\Delta \lambda=0.1$. However, we can see that the $p(f)$'s for these two cases are more focused than the one with $h_z=1.0$ (green pluses). This implies the tendency for the effect of  ground state degeneracy.

\section{Two numerical consistency checks }

In this Appendix, we demonstrate two consistency checks relating to work statistics numerically.

\subsection{Equivalence between thermal representations by MPO and METTS}\label{app_a1}

\begin{figure}
 \includegraphics[width=0.45\textwidth]{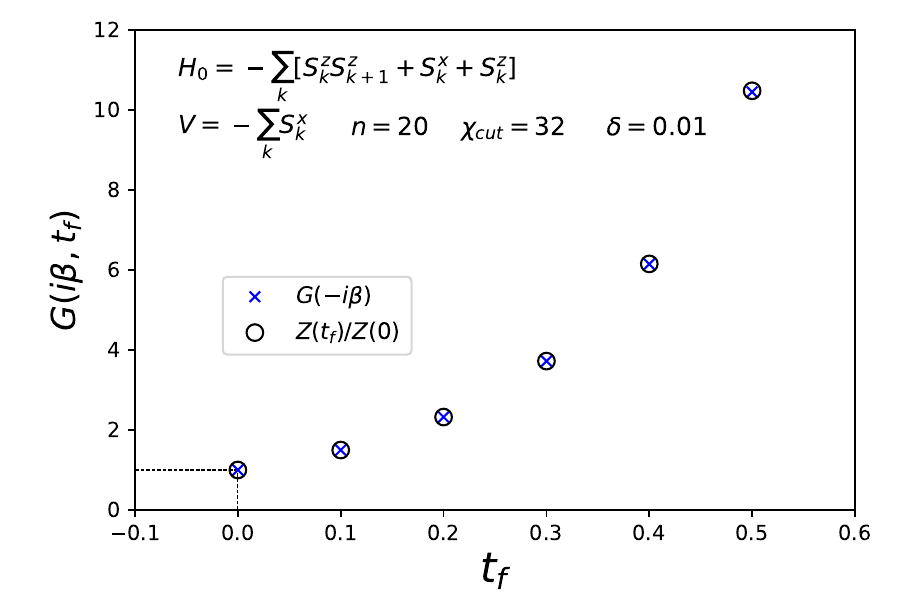}
 \caption{\small 
 Numerical agreement of the results for the Jarzynski equality. We evaluate work characteristic function $G(i\beta;t_f)\vert_{\beta=1}$ (crosses) and the ratio of partition functions ${Z(t_f) \over Z(0)}\Big\vert_{\beta=1}$ (circles) for $n=20$ quantum Ising chain by TEBD and MPO of $\chi_{cut}=32$ and $\delta=0.01$ to represent the thermal states.  The result shows Jarzynski equality holds and justifies using MPO for constructing the thermal states. After comparison, we also find that our results numerically agree with the ones from 
 the ensemble of $10000$ METTS, as shown in Fig. 4 in Ref.~\cite{gu2022tensor}.
}
 \label{fig:thermal_state}
\end{figure}

We here present in Fig.~\ref{fig:thermal_state} our numerical results of work characteristic function $G(i\beta;t_f)\vert_{\beta=1}$ and the ratio of partition functions ${Z_(t_f) \over Z(0)}|_{\beta=1}$ for checking the fluctuation theorem of $n=20$ quantum Ising chain by TEBD and the MPO representation for the thermal states. 
Our results agree numerically with the one from METTS, shown in Fig. 4 of Ref.~\cite{gu2022tensor}. 
This can justify our usage of MPO representation of thermal states for examining the passivity of thermal states, as shown in Fig.~\ref{fig:passivity_cIaing}. 

\subsection{Equivalence between average works done obtained from Eq.~\eq{work_realtime} or from  Eq.~\eq{w_sq}}  \label{app_a3}

\begin{figure}
\includegraphics[width=0.49\textwidth]{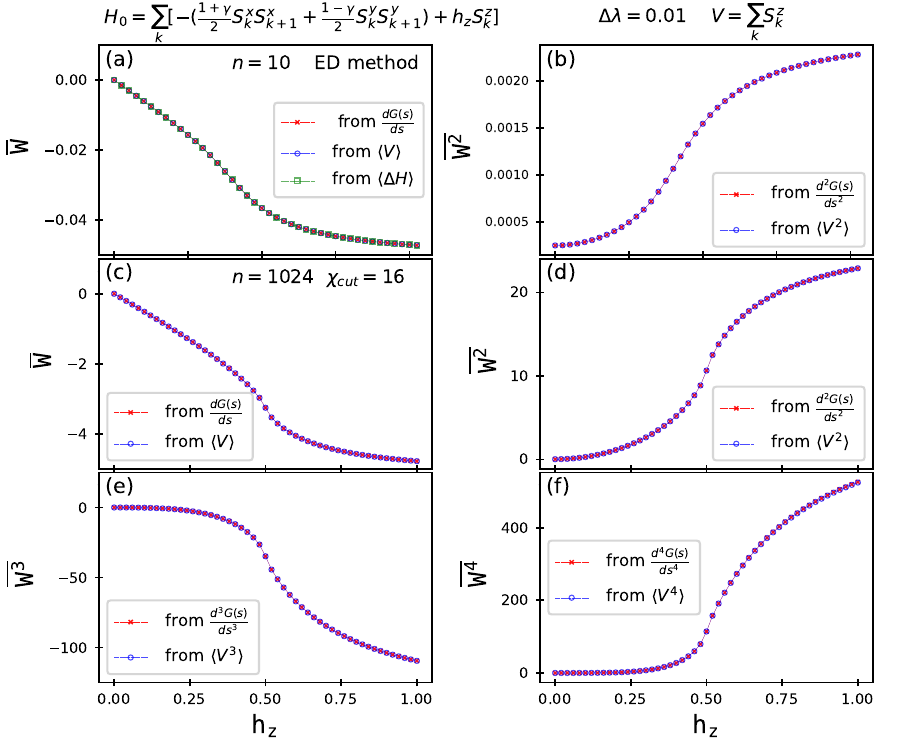}
\caption{\small  
Agreement of $\overline{W^m}$ of $m=1,2,3,4$ obtained from either characteristic function $G(-is)$ of  Eq.~\eq{work_realtime} (red crosses) or from the vev of $V^m$ (blue circles) for the transverse quantum Ising chains under sudden quench process with $V=\sum_k S_k^z$ and $\Delta \lambda=0.01$. In (a) and (b), the result for $n=10$ chains is obtained by ED method, and for (a) it also shows the result from the vev of Hamiltonian offset by Eq.~\eq{w_sq} (green squares). In (c),(d),(e), and (f), the results for $n=1024$ chains are calculated by the MPS-based method.  
}
 \label{fig:Ising_consistent_test}
\end{figure}

\begin{figure}
\includegraphics[width=0.49\textwidth]{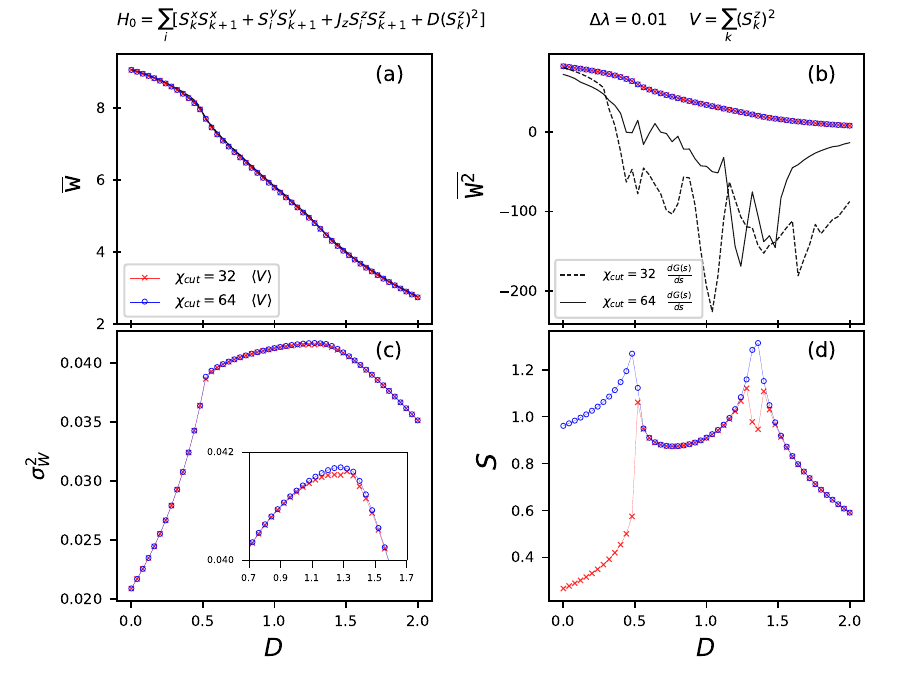}
\caption{\small  
Examining the agreement of (a) $\overline{W}$, (b) $\overline{W^2}$, (c) $\sigma^2_W$ for $n=1024$ Haldane (XXZ)  chain with various coupling $D$ in the initial Hamiltonian $H_0$ of size $n=1024$ under sudden quench process with $V=\sum_k (S_k^z)^2$ and $\Delta \lambda=0.01$. These results are obtained from the characteristic function $G(-is)$ (black dashed line for $\chi_{cut}=32$ or solid line for $\chi_{cut}=64$) or from the vev of $V^m$ (red crosses for $\chi_{cut}=64$ or blue circles $\chi_{cut}=64$). We see that 
the results from both ways match well for (a) and (c) but not for (b). We thus evaluate entanglement entropy $S$ in (d) to show the relevance of increasing the bond dimension $\chi_{cut}$ in improving the accuracy, which could also be the reason for the mismatch in (b). 
}
 \label{fig:XXZ_consistent_test}
\end{figure}

There are two ways to obtain the average work done by a sudden quench process. The first one is to extract it from the work characteristic function $G(u)$ of Eq.~\eq{work_realtime}  by using Eq.~\eq{w_avg}. The second one calculates it by evaluating the vacuum expectation value (vev) of the Hamiltonian offset using Eq.~\eq{w_sq}. We check the agreement of the results from both ways for the quantum Ising and Haldane chains of $n=10$ by ED or $n=1024$ by MPS-based method, as shown in Fig. \ref{fig:Ising_consistent_test} and Fig. \ref{fig:XXZ_consistent_test}, respectively.

In Fig.~\ref{fig:Ising_consistent_test} for the quantum Ising model with various $h_z$ in the initial Hamiltonian $H_0$ under sudden quench process with $V=\sum_k S_k^z$ and $\Delta \lambda=0.01$, we show that the numerical agreement of the numerical results of $\overline{W^m}$ for $m=1,2,3,4$ obtained from the characteristic function $G(-is)$ (red crosses) or from the vev of $V^m$ (blue circles). This serves as a consistency check for the TEBD method.

In Fig.~\ref{fig:XXZ_consistent_test} for $n=1024$ Haldane (XXZ)  chain with various coupling $D$ in the initial Hamiltonian $H_0$ under sudden quench process with $V=\sum_k (S_k^z)^2$ and $\Delta \lambda=0.01$, we show that results for (a) $\overline{W}$, (b) $\overline{W^2}$, (c) $\sigma^2_W$ and (d) entanglement entropy $S$, which are obtained from the characteristic function $G(-is)$ (black dashed line for $\chi_{cut}=32$ or solid line for $\chi_{cut}=64$) or from the vev of $V^m$ (red crosses for $\chi_{cut}=64$ or blue circles $\chi_{cut}=64$). We can see that the results from both ways match well for (a) and (c) but not for (b). This mismatch can be due to the need of large $\chi_{cut}$ to compensate for the errors of performing numerical derivatives on $G(-is)$. This is seen from the improvement of $\overline{W^2}$ in (b) and $S$ in (d) when enlarging $\chi_{cut}$ twice large.

\begin{figure*}
\includegraphics[width=1.0 \textwidth]{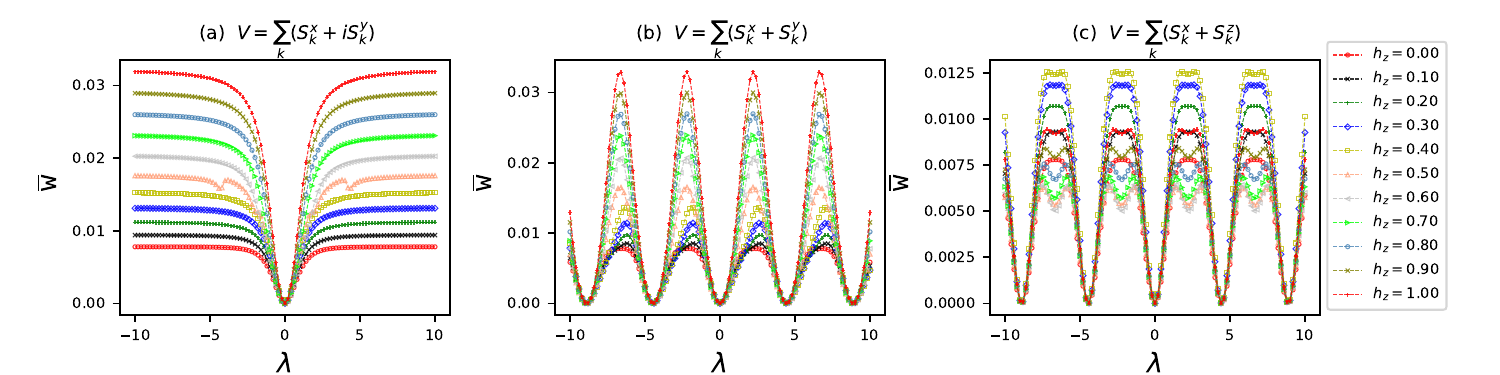}
 \caption{\small 
   Plot of $\overline{w}^{\rm impulse}$ to show the passivity of ground states and its pattern as a function of the coupling $\lambda$ of  $n=1024$ quantum Ising chain.  We fix $h_x=0$ but vary $h_z \in [0,1]$ to obtain various ground states, and also vary $\lambda \in [-10,10]$ under impulse process implemented by Hamiltonian of Eq.~\eq{H(t)} and Eq.~\eq{couple_impulse} with the non-Hermitian actions   
   (a) $V=\sum_k (S_k^x + i S_k^y)$, 
   (b) $V=\sum_k (S_k^y + S_k^y) $,and  
   (c) $ V= \sum_k (S_k^x + S_k^z)$. }
\label{fig:passivity_ising_v2}
\end{figure*}

\section{ More results for examining the passivity of ground states in Ising-like and Haldane-like chains} \label{App:passive}

In this Appendix, we present more results of examining the passivity of the ground states. The first one is shown in Fig.~\ref{fig:passivity_ising_v2} for the plot of $\overline{W}^{\rm impulse}$ of Ising-like chains as a function of $\lambda \in [-10,10]$ for the more general non-Hermitian actions. It can be seen as the non-Hermitian counterpart of Fig.~\ref{fig:passivity_ising}.  

\begin{figure}[ ]
\includegraphics[width=0.45\textwidth]{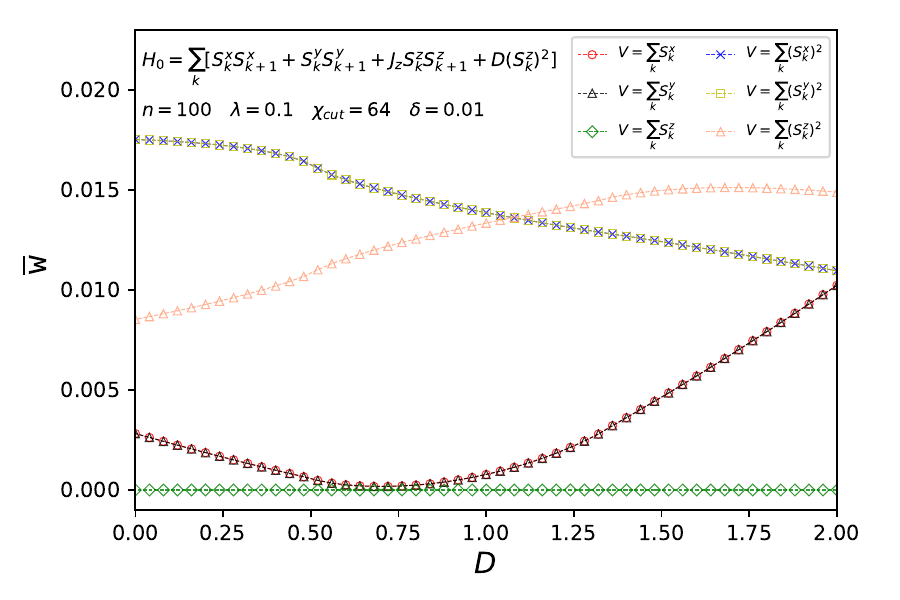}
 \caption{\small 
 Plot of $\overline{w}^{\rm impulse}$ to show the passivity of ground states of $n=1024$ Haldane-like chain with $J_z=1.5$ but varying $D \in [0,2]$ under impluse process of $\lambda=0.1$ and Hermitian actions  $V=\sum_k S_k^x$ (black triangles), $\sum_k S_k^y$ (yellow right-arrows), $\sum_k S_k^z$ (red circles), $\sum_k (S_k^x)^2$ (green diamonds), $\sum_k (S_k^y)^2$ (orange left-arrows) and $\sum_k (S_k^z)^2$ (blue squares). 
 The $\overline{W}^{\rm impluse}$ plots show that the ground states are always passive for all considered $V$'s. The degeneracy of $\overline{W}^{\rm impluse}$ under swapping of $S_i^x$ and $S_i^y$, as shown here, can be understood as the $Z_2$ symmetry of $H_0$ and $SO(3)$ spin algebra under the transformation: $(S_k^x,S_k^y,S_k^z)\rightarrow (-S_k^y,S_k^x,S_k^z)$.  Besides, $\overline{W}^{\rm impluse}$ always vanishes for $V=\sum_k S_k^z$ because of $[H_0,\sum_k S_k^z]=0$. }
\label{fig:passivity_QXXZ_W}
\end{figure}

\begin{figure*}[ ]
\includegraphics[width=1.0 \textwidth]{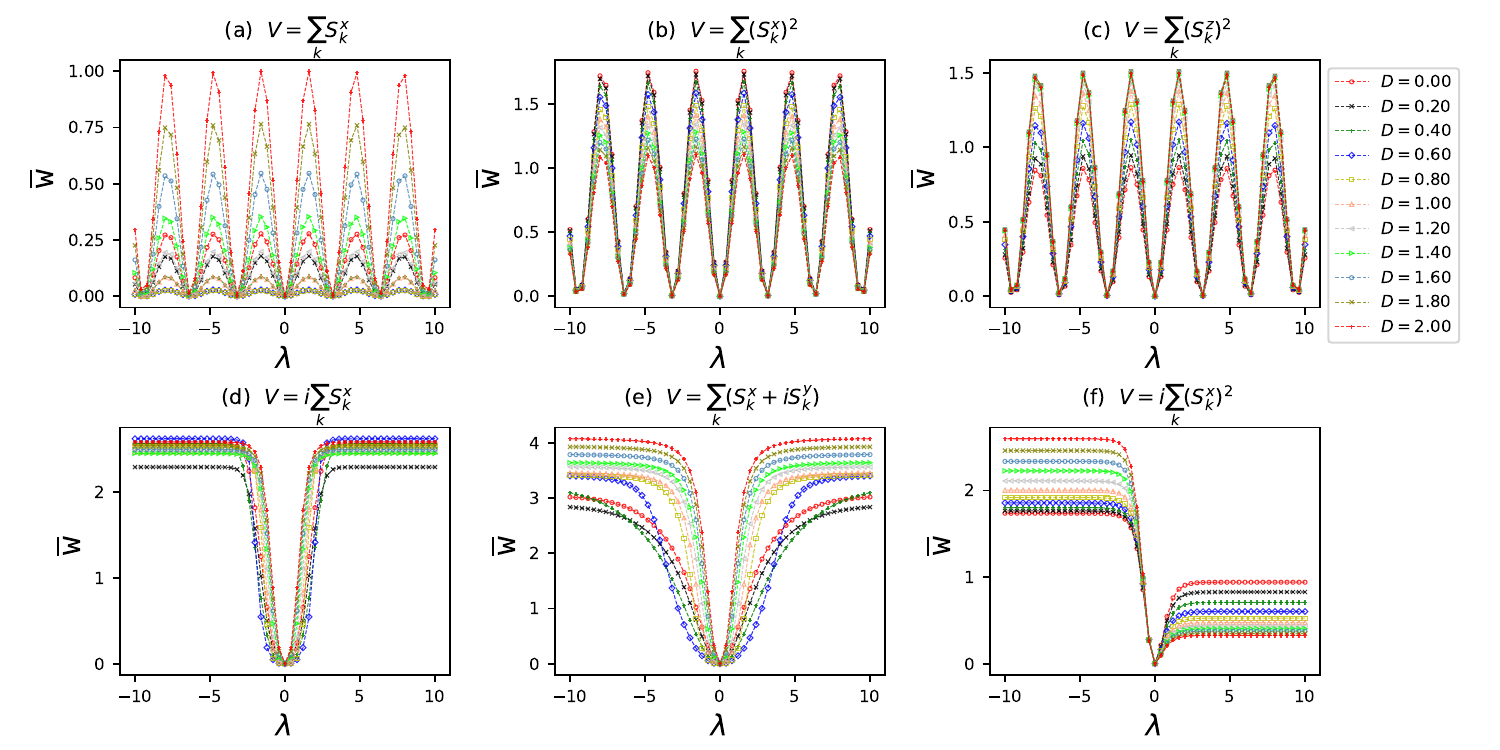}
 \caption{\small 
   Plot of $\overline{w}^{\rm impulse}$ to show the passivity of ground states and its pattern as a function of the coupling $\lambda$ of $n=1024$  Haldane-like chain with $J_z=1.5$.  We vary the coupling $D \in [0,2]$ to obtain various ground states, and also vary $\lambda \in [-10,10]$ under impulse process implemented by Hamiltonian of Eq.~\eq{H(t)} and Eq.~\eq{couple_impulse} for the Hermitian actions  (a) $V=\sum_k S_k^x$, (b) $V=\sum_k (S_k^x)^2$, (c) $ V= \sum_k (S_k^z)^2$ in the top row, and non-Hermitian actions (d) $V= i \sum_k S_k^x$, (e) $V= \sum_k (S_k^x + i S_k^y )$, and (f) $ V= i\sum_k ( S_k^x)^2$ in the bottom row. }
\label{fig:passivity_XXZ_all}
\end{figure*}

Then, we present the Haldane-chain counterparts of Fig.~\ref{fig:passivity_QIsing_W} and Fig.~\ref{fig:passivity_ising}, respectively, in Fig. \ref{fig:passivity_QXXZ_W} and Fig. \ref{fig:passivity_XXZ_all}.  Fig.~\ref{fig:passivity_QXXZ_W} shows the plot of $\overline{W}^{\rm impulse}$ as a function of coupling parameter $D$ labeling the ground states of the Haldane-like chains under various Hermitian actions for small value of $\lambda=0.1$. Fig.~\ref{fig:passivity_XXZ_all} shows the plot of $\overline{W}^{\rm impulse}$ as a function of $\lambda$ to examine the passivity of the ground states under either the Hermitian or non-Hermitian actions of a cyclic impulse process.  The key features are similar to what we have observed in Fig.~\ref{fig:passivity_QIsing_W} and Fig.~\ref{fig:passivity_ising} of the main text for the Ising chains. In particular, the ground states are all passive for all the cases considered.  Moreover, the features shown in  Fig.~\ref{fig:passivity_XXZ_all} are kind of the mixtures of the top and bottom rows in Fig.~\ref{fig:passivity_ising} as the $V$'s considered here are the mixtures of those twos.

\FloatBarrier

%


\end{document}